\documentclass{aa}

\usepackage{graphicx}
\usepackage{amsmath} 
\usepackage{amssymb}
\usepackage{lscape}
\usepackage{longtable,lscape}
\usepackage{aalongtable,lscape}
\usepackage[sort&compress]{natbib}
\usepackage{natbib}
\usepackage{pifont}

  \bibpunct{(}{)}{;}{a}{}{,}

\usepackage{txfonts}

\begin{document}

   \title{A deep multi-band investigation of IC~2391 
			\thanks{Based on observations carried out at the ESO/La Silla, 
			Chile under proposal number 68.D-0541(A).}}

  \author{L. Spezzi\inst{1,2} 
   \and I. Pagano\inst{1}
   \and G. Marino\inst{1}
   \and G. Leto\inst{1}
   \and E. Young\inst{3}
   \and N. Siegler\inst{4}
   \and Z. Balog\inst{3}
   \and S. Messina\inst{1}
   \and E. Distefano\inst{1}
   \and B. Mer\'in\inst{5}
   \and D. Barrado y Navascu\'es\inst{6}
   }

   \offprints{L. Spezzi, \email{lspezzi@rssd.esa.int}}

\institute{INAF - Osservatorio Astrofisico di Catania, via S. Sofia, 78, 95123 Catania, Italy 
\and European Space Agency (ESTEC), PO Box 299, 2200 AG Noordwijk, The Netherlands
\and Steward Observatory, University of Arizona, 933 North Cherry Avenue, Tucson, AZ 85721, USA
\and Jet Propulsion Laboratory, California Institute of Technology, Pasadena, California 91109-8099, USA 
\and European Space Agency (ESAC), PO Box Apdo. de correos 78, 28691 Villanueva de la Ca\~nada, Madrid, Spain
\and Laboratorio de Astrofísica Espacial y Exoplanetas, Centro de Astrobiologia (LAEFF-CAB, INTA-CSIC), 
European Space Astronomy Center (ESAC), PO Box 78, 28691 Villanueva de la Cañada, Madrid, Spain
}


   \date{Received ; accepted }

  \abstract
{}
{We report the outcome of a deep multi-wavelength study 
of the IC~2391 young open cluster. We aim at uncovering 
new low-mass and sub-stellar members of the cluster 
and identifying new debris disk objects.}
{A 30$\times$30~square arcmin area in IC~2391 was 
observed using the wide-field imager at the 
ESO 2.2m telescope. The completeness limits of the photometry 
at 3$\sigma$ level are $V$=24.7, $R_C$=23.7 and $I_C$=23.0, faint enough 
to reveal sub-stellar members down to $\sim$0.03~M$_\odot$. 
Our membership criteria are based on the use of our optical data, in combination 
with $JHK_S$ magnitudes from the 2MASS catalog. 
We also estimate the physical parameters of the selected 
candidates. Debris disk candidates 
are identified on the basis of their infrared excess emission using 
near- and mid-infrared photometry from the Spitzer Space Telescope.}
{Our optical survey, which has a limiting magnitude at 3$\sigma$ 
level 1-2~mag fainter than 
previous optical surveys conducted in IC~2391, revealed 
29 new low-mass member candidates of the cluster. 
We estimate the contamination to be at least $\sim$50\%. 
We constrain the fraction of sub-stellar objects in the range 8-15\% 
and discuss possible explanations for the deficit of brown dwarfs in this cluster. 
We also identified 10 candidates in the cluster 
showing IR excess emission consistent with the presence of debris disks.}
{}

   \keywords{stars: low-mass, brown dwarfs -- stars: formation -- stars: pre-main sequence -- 
			stars: circumstellar matter -- ISM: individual objects: IC~2391}

   \maketitle

\section{Introduction \label{intro}}

Over the last decade the general form of the stellar IMF has been 
observationally established, while the situation remains less 
solid across the hydrogen-burning limit \citep{Rei05}. More recently, 
infrared (IR) survey data \citep[e.g.][]{Bou98,Mor03,Mue03,Mue07,Luh04b,Luh06,Lod05b,Lod06a,Lod06b,Lod06c,Lod07}, 
have produced reliable statistics for substellar objects. 
These overall results, in combination with optical photometric and 
spectroscopic follow-up observations, allow to quantitatively address 
such issues as the continuity of the IMF across the stellar/substellar boundary, 
its invariance or dependency on local conditions, etc. 
Such studies also hold important keys for the clarification 
of the dominant brown dwarf (BD) formation mechanism in star clusters \citep{Bou05}.	

Because of its proximity \citep[$\sim$150~pc;][]{For01}, relatively young age 
\citep[30-50~Myr;][]{Mer81,Bas96,Sta98,Bar99,Bar04} and insignificant 
extinction along the line of sight \citep[E(B-V)=0.006;][]{Pat96}, 
the IC~2391 young open cluster is long since considered an ideal location 
for star formation studies. 
Consequently, IC~2391 is one of the best studied clusters. 
There are about 180 members known in this cluster, with accurate 
proper motion determinations \citep{Mon03,Dod04,Pla07}, 
among which a dozen of substellar objects. 
The only determination of the IC~2391 mass spectrum in the very low-mass regime has been presented by \citet{Bar04}. 
According to these authors, the mass spectrum between 0.5~M$_\odot$ and the substellar limit 
($\sim$0.07~M$_\odot$) follows a power law with $\alpha$-index $\sim$0.96. 
Below this limit, there is a sudden drop that the authors attribute to a 
local drop in the shape of the luminosity-mass relation at spectral types M7-M8 and, 
partially, to the incompleteness of their spectroscopic follow-up below $I_C$=18.5~mag 
(i.e. 0.05~M$_\odot$ for cluster members).

However, other young clusters (e.g. $\alpha$~Per and NGC~2547) 
show a lack of substellar members more or
less with the same masses as in the case of IC~2391 \citep[see][]{Bar02,Jef04}. 
This deficit may extend to lower masses or may only be a dip, 
perhaps caused by an imperfect understanding of the mass-luminosity 
relationship at temperatures around 2400-2500~K \citep{Dob02}. However, 
the incompleteness of the surveys conducted in these clusters leaves the question open.
In particular, the number of substellar objects discovered so far in IC~2391 
is too low to draw any conclusion on a statistically significant basis.

In this paper, we present the results of a deep optical, wide-field imaging survey 
of IC~2391, complemented by IR photometry from the 2MASS catalog \citep{Cut03} 
and the Spitzer satellite. 
This survey, which has a 3$\sigma$ photometric limit 1-2 magnitudes fainter 
than previous optical photometric surveys conducted in IC~2391 
\citep[i.e. $I_C <$21-22~mag;][]{Pat96,Pat99,Bar01b}, 
aims at uncovering new low-mass and BD member candidates, 
to be confirmed with future follow-up spectroscopy, in order to provide 
further insights into the substellar mass spectrum in this cluster. 
In addition, the age and distance of IC~2391 offer a unique combination to 
study the evolution of debris disks around low-mass stars, which is 
a crucial step toward our understanding of 
the terrestrial planet formation mechanism \citep[e.g.][]{Hab01,Dec03,Dom03,Cur07}. 
The study of circumstellar disks has recently undergone a 
substantial improvement thanks to the data from the Spitzer Space Telescope. 
These data are very adequate for disk investigations because 
the Spitzer sensitivity and wavelength coverage allow 
studies of statistically significant samples and
probe the inner planet-forming region of disks \citep[see, e.g.,][]{Lad06}. 
Spitzer observations have shown that debris disks are found around solar-like 
stars at a wide range of possible distances 
(1-50~AU) from the central star and temperatures (10-650K), 
with an age-dependent frequency \citep[see][ and reference therein]{Sie07}. 
A pioneer study of the debris disk population in IC~2391 has been 
recently conducted by \citet{Sie07} using the Multiband Imaging 
Photometer for Spitzer (MIPS). 
Using our optical photometry. the MIPS catalog by \citet{Sie07} 
and additional IR photometry from 2MASS and the 
InfraRed Array Camera (IRAC) on Spitzer, we search for 
new debris disk candidates in IC~2391. 
These studies ultimately aim at measuring the frequency of debris disks around very low-mass stars, 
in order to assess whether they experience a planetesimal phase in their evolution, 
as observed for more massive stars \citep[e.g.][]{You04,Che05,Lad06,Ria06,Kes06,Sic07}.

In Sect.~\ref{sec_obs} we describe the WFI survey in IC~2391 
and the data reduction and catalog extraction procedures.
In Sect.~\ref{selection} and Sect.~\ref{contamination} we describe our membership criteria, 
present the sample of new cluster member candidates and estimate 
the contamination level of this sample. 
The fraction of sub-stellar objects in IC~2391 is discussed in Sect.~\ref{IMF_BD}, while in 
Sect.~\ref{sec_debris} we investigate the debris disk population in this cluster.
Our conclusions are drawn in Sect.~\ref{conclu}.

\section{Observations and data reduction\label{sec_obs}}

\subsection{The WFI observations \label{obs}}

\begin{figure*} 
\centering
\includegraphics[width=16cm,height=14cm]{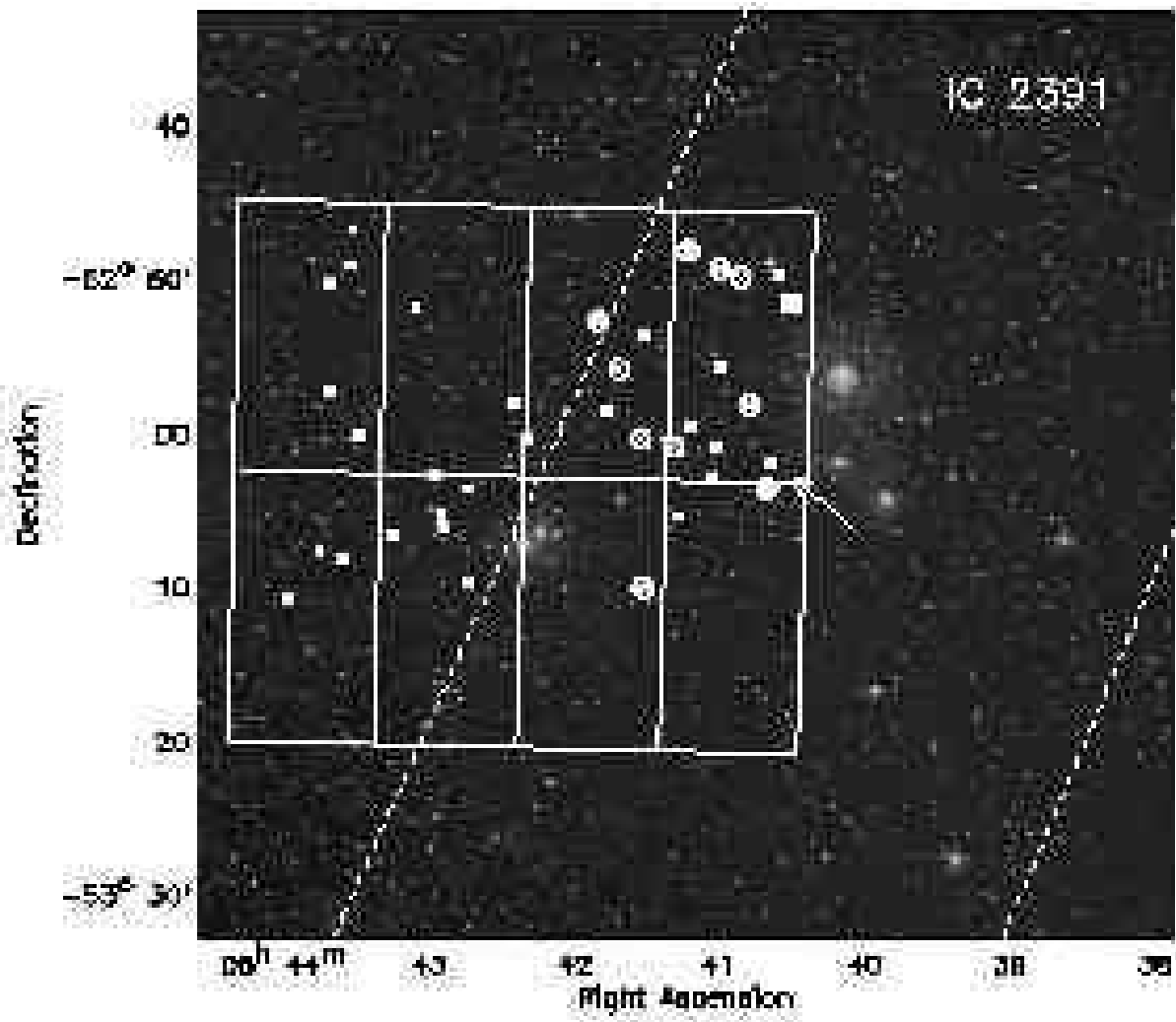}
\caption{$R$-band image of IC~2391 from the STScI Digitised Sky Survey. 
The solid line display the total sky-area surveyed with the WFI 
CCD mosaic camera at the ESO 2.2m telescope while the dashed lines trace 
the area observed with IRAC and MIPS at Spitzer (see Sect.~\ref{sec_debris}). 
The arrow indicates the cluster center The filled squares represent 
the new member candidates identified in this work, while the circled 
symbols indicate the objects showing IR excess emission (the stars 
mark the confirmed cluster members, while squares and 
crosses mark the member candidates and the four sources of dubious nature 
discussed in Sect.~\ref{note_obj}, respectively).}
\label{fig:obs}
\end{figure*}

The optical photometric observations were performed in service mode using 
the Wide Field Imaging (WFI) mosaic camera attached 
to the ESO 2.2m telescope at La Silla (Chile), from November 
2001 to March 2002 (Program ID: 68.D-0541(A), PI: Isabella Pagano). 
WFI is an $8k \times 8k$ CCD mosaic, covering a 
30$\times$30~square arcmin field with a pixel scale 
of 0.238 arcsec/pix. 
These observations are also intended to investigate the variability of 
young stars at the bottom of the main sequence due to their magnetic activity, 
which will be presented in a future paper (Pagano et al., in preparation). 
For this reason, our observing strategy consisted of 
repeated exposures of the same field. 
In order to avoid photometric contamination from the 
bright AB type stars in the core of IC~2391, our WFI pointing is 
centered at R.A.=08$^h$:42$^m$:31$^s$ and Dec.=$-$53$^o$:02$^m$:58$^s$, 
i.e. about 0.25~deg east from the cluster center (Fig.~\ref{fig:obs}), and 
covers about 60\% of the sky-area spanned by IC~2391. 
The region was observed in the $V$, $R$ and $I$ WFI broad-band filters 
for nine non-consecutive nights, with 1-7 observations per night per filter. 
The seeing conditions were always better than 1.2~arcsec.
This strategy ensured a good phase coverage for objects 
with rotational periods between 0.2 and 10 days, i.e. 
the range expected for very low-mass stars at the age of IC~2391 (30-50~Myr). 
In order to cover the gaps between the WFI CCDs, each observation in
each filter was split into three individual exposures (ditherings), 
shifting the telescope pointing by $\sim$1~arcmin 
between consecutive exposures. 
Since we are interested both in identifying the lowest mass 
cluster members and recovering the early type ones, 
we performed two series of observations per night. 
The total exposure time was 270~s in
the $V$-band and 75~s in the $RI$ bands 
for the ``long-time'' series and 30~s in the $V$-band, 15~s in 
the $R$-band and 21~s in the $I$-band for the ``short-time'' one.
In this way, the photometry for early type members saturated 
in the long-time exposures is recovered from the ``short-time'' ones. 
A summary of the observations is reported in Table~\ref{tab:obs}.

\begin{table*}
\caption[ ]{\label{tab:obs} Journal of the observations.}
\begin{center}
\scriptsize
\begin{tabular}{clclc|clclc}
\hline
 Date       &  WFI Filter  & MJD & T$_{\rm exp}$ &  Air Mass  & Date       &  WFI Filter  & MJD & T$_{\rm exp}$ & Air Mass \\
 (d/m/y)    &              &     &   (sec)       &            & (d/m/y)    &              &     &   (sec)       &          \\
\noalign{\medskip} \hline
24/11/2001 &  V/89 	&	52237.22128823 &	3$\times$10 &	1.50  & 	04/03/2002 &  V/89  &	52337.20706880 &	3$\times$10 &	1.22  \\	
           & Rc/162 & 52237.22634835 & 3$\times$5  & 1.47  &            	& Rc/162 & 52337.21803382 & 3$\times$5  & 1.25  \\ 
           & Ic/Iwp & 52237.23117713 & 3$\times$7  &	1.44  &            	& Rc/162 & 52337.22950399 & 3$\times$25 &	1.29  \\ 
          	& Ic/Iwp & 52237.23529559	& 3$\times$25 &	1.42  & 	           & 	V/89  &	52337.23437902 &	3$\times$90 &	1.31  \\	
          	& Rc/162 & 52237.24074915 & 3$\times$25 &	1.39  &             &  V/89  &	52337.24109513 &	3$\times$10 &	1.34  \\	
\cline{6-10}																																																					
	          & 	V/89 	&	52237.24882797 &	3$\times$90 &	1.35  &  05/03/2002 &  V/89  &	52338.21350087 &	3$\times$10 & 1.24  \\	
											&  V/89 	&	52237.26303437 &	3$\times$10 &	1.30  &            	& Rc/162 & 52338.21877655 & 3$\times$5  & 1.26  \\ 
          	& Rc/162 & 52237.26868166 & 3$\times$5  & 1.28  &            	& Ic/Iwp & 52338.22356216	& 3$\times$7  & 1.28  \\ 
          	& Ic/Iwp & 52237.27821246	& 3$\times$7  & 1.24  &            	& Ic/Iwp & 52338.22763522	& 3$\times$25 & 1.29  \\	
          	& Ic/Iwp & 52237.28227565	& 3$\times$25 &	1.23  &             & 	V/89  &	52338.23909508 &	3$\times$90 & 1.34  \\	
\cline{6-10}																																																					
          	& Rc/162 & 52237.28773647 & 3$\times$25 &	1.21  &  06/03/2002 &  V/89  &	52339.18778127 &	3$\times$10 &	1.18  \\	  
           & 	V/89 	&	52237.29344333 &	3$\times$90 &	1.20  &            	& Rc/162 & 52339.19297910 & 3$\times$5  & 1.19  \\ 
           &  V/89 	&	52237.30066930 &	3$\times$10 &	1.18  &            	& Ic/Iwp & 52339.20270214	& 3$\times$7  & 1.22  \\  
          	& Rc/162 & 52237.31212718 & 3$\times$5  & 1.16  &             & Ic/Iwp & 52339.20714999	& 3$\times$25 &	1.23  \\	
          	& Ic/Iwp & 52237.31721340	& 3$\times$7  & 1.15  &            	& Rc/162 & 52339.21242581 & 3$\times$25 &	1.25  \\ 
\cline{1-5}                                                 
26/11/2001 &  V/89 	&	52239.18065529 &	3$\times$10 &	1.78  &             & 	V/89  &	52339.21805674 &	3$\times$90 &	1.27  \\	  
          	& Rc/162 & 52239.18562202 & 3$\times$5  & 1.74  &             &  V/89  &	52339.22604426 &	3$\times$10 &	1.30  \\	
          	& Ic/Iwp & 52239.19029892	& 3$\times$7  & 1.69  &            	& Rc/162 & 52339.23099193 & 3$\times$5  & 1.32  \\ 
          	& Ic/Iwp & 52239.19447743	& 3$\times$25 &	1.66  &            	& Ic/Iwp & 52339.23578713	& 3$\times$7  & 1.34  \\ 
          	& Rc/162 & 52239.19983479 & 3$\times$25 &	1.62  &            	& Ic/Iwp & 52339.23999762	& 3$\times$25 &	1.35  \\	
           & 	V/89 	&	52239.20535873 &	3$\times$90 &	1.57  &            	& Rc/162 & 52339.24539390 & 3$\times$25 &	1.38  \\ 
           &  V/89 	&	52239.32447597 &	3$\times$10 &	1.13  &             & 	V/89  &	52339.25120870 &	3$\times$90 &	1.41  \\	
          	& Rc/162 & 52239.32957951 & 3$\times$5  & 1.12  &             &  V/89  &	52339.25913018 &	3$\times$10 &	1.45  \\	
          	& Ic/Iwp & 52239.33441270	& 3$\times$7  & 1.12  &             & Rc/162 & 52339.26466470 & 3$\times$5  & 1.49  \\ 
          	& Ic/Iwp & 52239.33845778	& 3$\times$25 &	1.11  &            	& Ic/Iwp & 52339.26890995	& 3$\times$7  & 1.51  \\ 
          	& Rc/162 & 52239.34398065 & 3$\times$25 &	1.11  &            	& Ic/Iwp & 52339.27256372	& 3$\times$25 &	1.53  \\	
           & 	V/89 	&	52239.35053232 & 3$\times$90 &	1.10  &             & Rc/162 & 52339.28297103 & 3$\times$25 &	1.61  \\ 
\cline{1-5}	                                                
23/01/2002 &  V/89  &	52297.19774832 &	3$\times$10 &	1.10  &             & 	V/89  &	52339.28856480 &	3$\times$90 &	1.66  \\	
          	& Rc/162 & 52297.20344695 & 3$\times$5  & 1.10  &             &  V/89  &	52339.29614328 &	3$\times$10 &	1.73  \\	
          	& Ic/Iwp & 52297.20807586	& 3$\times$7  & 1.09  &            	& Rc/162 & 52339.30145210 & 3$\times$5  & 1.78  \\ 
          	& Ic/Iwp & 52297.21229543	& 3$\times$25 &	1.09  &            	& Ic/Iwp & 52339.30606411	& 3$\times$7  & 1.82  \\ 
          	& Rc/162 & 52297.21764270 & 3$\times$25 &	1.09  &            	& Rc/162 & 52339.31534623 & 3$\times$25 &	1.93  \\ 
\cline{6-10}	                                                
           & 	V/89  &	52297.22325254 &	3$\times$90 &	1.09  &  24/03/2002 &  V/89  &	52357.07361738 &	3$\times$10 &	1.10  \\	
           &  V/89  &	52297.23089695 &	3$\times$10 &	1.09  &            	& Rc/162 & 52357.07891852 & 3$\times$5  & 1.10  \\ 
          	& Ic/Iwp & 52297.24111660	& 3$\times$7  & 1.10  &            	& Ic/Iwp & 52357.08412627	& 3$\times$7  & 1.10  \\ 
          	& Ic/Iwp & 52297.24521964	& 3$\times$25 &	1.10  &            	& Ic/Iwp & 52357.08840132	& 3$\times$25 &	1.10  \\	
          	& Rc/162 & 52297.25143005 & 3$\times$25 &	1.10  &            	& Rc/162 & 52357.09377871 & 3$\times$25 &	1.11  \\ 
           & 	V/89  &	52297.25701984 &	3$\times$90 &	1.11  &             & 	V/89  &	52357.09933803 &	3$\times$90 &	1.12  \\	
\cline{1-5}	 	                                              
11/02/2002 &  V/89  &	52316.14784279 &	3$\times$10 &	1.10  &             &  V/89  &	52357.10984608 &	3$\times$10 &	1.13  \\	
          	& Rc/162 & 52316.15586182 & 3$\times$5  & 1.09  &            	& Rc/162 & 52357.12092981 & 3$\times$5  & 1.15  \\ 
          	& Ic/Iwp & 52316.16066498	& 3$\times$7  & 1.09  &            	& Ic/Iwp & 52357.12572673	& 3$\times$7  & 1.16  \\ 
          	& Ic/Iwp & 52316.16510996	& 3$\times$25 &	1.09  &            	& Ic/Iwp & 52357.13007714	& 3$\times$25 &	1.16  \\	
          	& Rc/162 & 52316.17086954 & 3$\times$25 &	1.09  &            	& Rc/162 & 52357.13532178 & 3$\times$25 &	1.17  \\ 
           &  V/89  &	52316.17698296 &	3$\times$90 &	1.09  &             &  V/89  &	52357.16004606 &	3$\times$10 &	1.24  \\	
           &  V/89  &	52316.18480783 &	3$\times$10 &	1.10  &            	& Rc/162 & 52357.16513120 & 3$\times$5  & 1.26  \\ 
          	& Rc/162 & 52316.19440469 & 3$\times$5  & 1.10  &            	& Ic/Iwp & 52357.16995618	& 3$\times$7  & 1.27  \\ 
          	& Ic/Iwp & 52316.19919322	& 3$\times$7  & 1.10  &            	& Ic/Iwp & 52357.17390633	& 3$\times$25 &	1.28  \\	
          	& Ic/Iwp & 52316.20328731	& 3$\times$25 &	1.11  &            	& Rc/162 & 52357.18004707 & 3$\times$25 &	1.31  \\ 
          	& Rc/162 & 52316.20891675 & 3$\times$25 &	1.11  &             & 	V/89  &	52357.18570452 &	3$\times$90 &	1.33  \\	
           & 	V/89  &	52316.21545996 &	3$\times$90 &	1.12  &             & 	V/89  &	52357.99572072 &	3$\times$10 &	1.13  \\	
\cline{1-10}	                                                     
24/02/2002 &  V/89  &	52329.19058819 &	3$\times$10 &	1.14  &  25/03/2002	& Rc/162 & 52358.00131936 & 3$\times$5  & 1.13  \\ 
          	& Rc/162 & 52329.19609543 & 3$\times$5  & 1.14  &            	& Ic/Iwp & 52358.00615478	& 3$\times$7  & 1.12  \\ 
           & Ic/Iwp & 52329.20090155	& 3$\times$7  & 1.15  &            	& Rc/162 & 52358.01631727 & 3$\times$25 &	1.11  \\ 
          	& Rc/162 & 52329.21181380 & 3$\times$25 &	1.17  &             & 	V/89  &	52358.02278763 &	3$\times$90 & 1.10  \\	
           & 	V/89  &	52329.21777151 &	3$\times$90 &	1.19  &             &        &                &             &       \\
\noalign{\medskip} \hline                                     
\end{tabular}                                                 
\end{center}                                                 
\end{table*}

\subsubsection{Pre-reduction \label{prered}}

The raw images were processed using the IRAF\footnote{IRAF is 
distributed by NOAO, which is operated by the Association 
of Universities for Research in Astronomy, Inc., under 
contract to the National Science Foundation.} \emph{mscred} package 
and a number of scripts \emph{ad hoc} developed both under 
IRAF and under IDL\footnote{Interactive Data Language.}. 
We followed the standard steps for mosaic CCD data reduction, i.e. 
overscan, bias and flat-fielding correction. 
For each night in which observations for our program were performed, 
bias and twilight flat frames were combined to obtain the  
night master bias and flat, respectively, which were 
then used to correct the science images. 
Possible residual effects due to vignetting and sky concentration caused by the geometrical distortion 
were reduced by subtracting from each image a background's 2D-polynomial fit, 
obtained by using the \emph{imsurfit} routine under IRAF. 
Fringing strongly affects the WFI observations taken at near-IR 
wavelengths. In order to remove it from our $I$-band 
images, we subtracted from each $I$-band science frame 
the fringing pattern frame available as part 
of the WFI standard calibration plan\footnote{See: http://www.ls.eso.org/
lasilla/sciops/2p2/E2p2M/WFI/CalPlan/fringing/}, 
scaled by a specific factor to account
for the amplitude of the fringes in the individual science frames.

\subsubsection{Astrometry and co-addition of images \label{astrom}}

The astrometric calibration of each exposure was 
performed using the \emph{msczero} and \emph{msccmatch} 
packages under IRAF; the GSC \citep{GSC06} and 
TYCHO \citep{Hog98} catalogs were used as reference and an absolute 
astrometric precision of 0.6~arcsec was achieved, 
as confirmed by a cross-check with the 2MASS point
source catalog (see Sect.~\ref{catalog}). 
Individual exposures in each dithering set 
were then resampled, flux-scaled and combined into a single mosaic 
by using the \emph{mscimage}, \emph{mscimatch} and \emph{imcombine} 
tasks under IRAF, respectively. 
The final stacked image is a $8k \times 8k$ frame 
where each pixel value is the median flux of the 
co-added ditherings normalized to the total exposure 
time and relative to the airmass and atmospheric 
transparency of the first frame in the dithering set.
The final stacked images were also trimmed to the overlapping area
of the relative ditherings, so that only those sources falling 
in all ditherings are considered for the subsequent analysis.

\subsubsection{Photometric calibration\label{phot_cal}}

Instrumental magnitudes were reported to the 
standard Johnson-Cousins system. 
To this aim, the Landolt's standard star field SA~92 
\citep{Lan92} was observed in the $VRI$ filters.
By using the IRAF \emph{photcal} package, 
we first performed the aperture photometry for the standard stars, 
obtaining their instrumental magnitudes ($v_0$, $r_0$ and $i_0$) 
corrected for atmospheric extinction and normalized to the exposure time.
Then, the transformation coefficients, namely zero point ($ZP$) 
and color term ($c$), from the WFI photometric 
system to the Johnson-Cousins standard system 
were determined by a linear fitting of 
the following equations:
\begin{eqnarray}
V=v_0 + c_V \cdot (v_0-r_0)+ZP_V &\\
R_C=r_0 + c_R \cdot (r_0-i_0)+ZP_R &\\
I_C=i_0 +c_I \cdot (r_0-i_0)+ZP_I
\label{cal_VRI}
\end{eqnarray}
where $V$, $ R_C$ and $I_C$ are the standard magnitudes of the Landolt's stars. 
The mean transformation coefficients determined in our observing run 
are reported in Table~\ref{tab:coeff} and are consistent with 
the mean values computed for La Silla\footnote{See http://www.ls.eso.org/lasilla/Telescopes/2p2T/E2p2M/WFI/zeropoints/.}. 

\begin{table}
\caption[ ]{\label{tab:coeff} Mean photometric calibration coefficients for our WFI observing run.}
\begin{center}
\small
\begin{tabular}{cccc}  
\hline
Filter & {\it AE}$^\dagger$ & {\it ZP} & {\it c} \\
\noalign{\medskip}
\hline
\noalign{\medskip}
$V$		&  0.130 & 24.195$\pm$0.035 & 0.071$\pm$0.037  \\
$R$		&  0.096 & 24.509$\pm$0.027 & 0.078$\pm$0.031  \\
$I$		&  0.082 & 23.502$\pm$0.034 & 0.325$\pm$0.044  \\
\noalign{\medskip}
\hline
\end{tabular}
\end{center}
$^\dagger$ \footnotesize{AE = Atmospheric extinction coefficient. The mean values for La Silla have been adopted.}
\end{table}

\subsubsection{The deep catalog extraction \label{catalog}}

The sources extraction and photometry from 
each stacked image in each filter was performed 
using the SExtractor tool by \citet{Ber96}. 
This data set will be used to study the variability of the IC~2391 
low-mass members and will be presented in a future paper (Pagano et al., in preparation). 
The limiting magnitude achieved at the 3$\sigma$ level  
using the single staked images is $R \approx$20. 
At the age ($\sim$50~Myr) and distance ($\sim$150~pc) to IC~2391, 
this limit corresponds to very low-mass stars (M$\approx$0.1~M$_\odot$) 
close to the hydrogen burning limits or, perhaps, to massive BDs.

\begin{figure} 
\centering
\includegraphics[width=6cm,height=8cm]{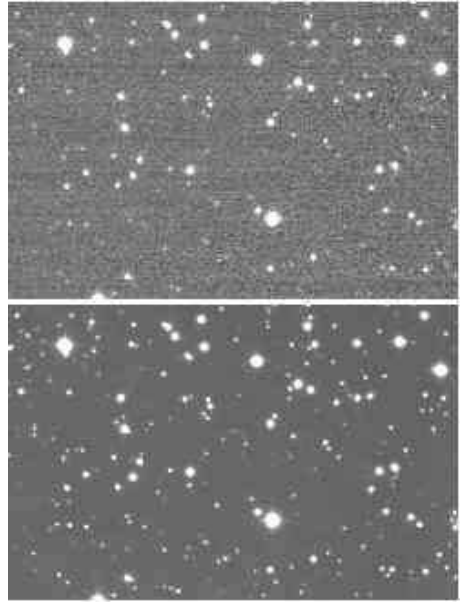}
\caption{Section of a single stacked image in the $R$-band
(upper panel) and the same region as it appears 
in the deeper frame obtained by combining 18 
stacked images (lower panel).}
\label{fig:deep}
\end{figure}

\begin{figure} 
\centering
\includegraphics[width=7cm,height=8cm]{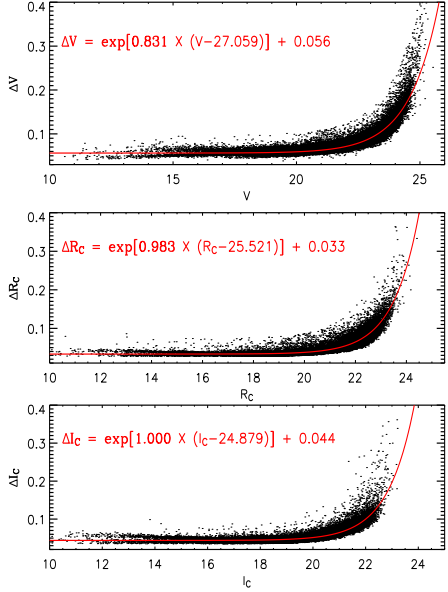}
\caption{Photometric errors versus magnitudes and relative 
exponential fits for all the point-like sources detected in 
the 30$\times$30~square arcmin surveyed area in IC~2391. 
Magnitudes and errors are from the deep frames obtained 
as discussed in Sect.~\ref{catalog}.}
\label{fig:err}
\end{figure}

One of the main goals of this work is the identification 
of new candidate substellar members of IC~2391, which are expected 
to have $R$ magnitude in the range 18.5-24 depending on their 
mass and the intervening reddening. 
Since our data set consists of repeated 
observations of the same field, we further combined the
stacked images with the same exposure time and 
obtained, for each filter, a photometrically deeper frame 
with improved signal-to-noise ratio, suitable 
for our purposes (Fig.~\ref{fig:deep}).
The combination has been performed using the \emph{imcombine} 
task under IRAF in the same configuration as explained in 
Sect.~\ref{astrom}, and the stacked images from our ``long-time'' 
series (see Sect.~\ref{obs}). 17 images were combined in the $V$ filter, 
18 in the $R$ filter and 15 in the $I$ filter, for a total 
exposure time of 76.5~min, 22.5~min and 18.75~min, respectively. 
PSF-fitting photometry was performed on these deeper frames 
by using the IRAF \emph{daophot} package \citep{Ste87}; 
this method allows a better deblending of sources with respect 
to aperture photometry and hence, the detection of the 
faintest neighbors to bright stars. 
Since we are interested in very low-luminosity objects, 
we set a very low detection threshold ($\sigma$=3) and 
then selected point-like objects from extended sources 
(e.g. galaxies), saturated objects, and other spurious detections 
by using the \emph{daophot} morphological parameters. 
In Fig.~\ref{fig:err} the internal photometric errors of 
all the detected point-like sources are plotted against 
magnitude for all the available filters; the relative 
exponential fits are over-plotted. 
Table~\ref{tab:mag_limits} reports the saturation limits and 
the limiting magnitudes achieved at the 3$\sigma$ level from these deeper frames.  
These limits are sufficient for the detection of 
young stars and BDs with mass from $\sim$1~M$_\odot$ down 
to $\sim$0.03~M$_\odot$ at the age and distance of IC~2391.

Our optical catalog contains 60596 point-like sources with $VRI$ magnitudes. 
A merged catalog with position and photometry in
each optical filter, as well as in the near-IR 
\citep[$JHK_S$ magnitudes from 2MASS;][]{Cut03}, was finally produced 
and used for the selection of cluster members 
(Sect.~\ref{selection}). A matching radius of 1\arcsec was used to merge the
WFI and the 2MASS catalogs; this value was set by taking 
the astrometric accuracy of both catalogs into account. 

\begin{table}
\caption[ ]{\label{tab:mag_limits} Saturation limits and limiting
magnitudes at 3$\sigma$ level (see Sect.~\ref{catalog}). 
In the fourth column the corresponding limiting mass for cluster members 
(i.e. assuming an age between 30 and 50~Myr and a distance of 150~pc) 
derived from the evolutionary models by \citet{Bar98} and \citet{Cha00} is reported; 
in the last column the magnitude corresponding to the Hydrogen burning limit 
($\sim$0.07~M$_{\odot}$) for IC~2391 members is also given for comparison.}
\begin{center}
\begin{tabular}{ccccc}
\hline
    Filter &  Mag Sat. &  Mag 3$\sigma$ & Mass 3$\sigma$ & PHB$^\dagger$\\
           &           &                &  (M$_\odot$)   &              \\
\noalign{\medskip} \hline
  $V$	  & 10.5 & 24.7 & 0.03  & 20.0 \\
  $R_C$   & 10.0 & 23.7 & 0.03  & 18.5 \\
  $I_C$   &  9.5 & 23.0 & 0.02  & 17.0 \\
\noalign{\medskip} \hline
\end{tabular}
\end{center}
$^\dagger$ \footnotesize{PHB = Photometric Hydrogen Burning limit for IC~2391 members.}
\end{table}

\subsubsection{Completeness \label{psf_compl}}
 
\begin{figure} 
\centering
\includegraphics[width=9cm,height=7cm]{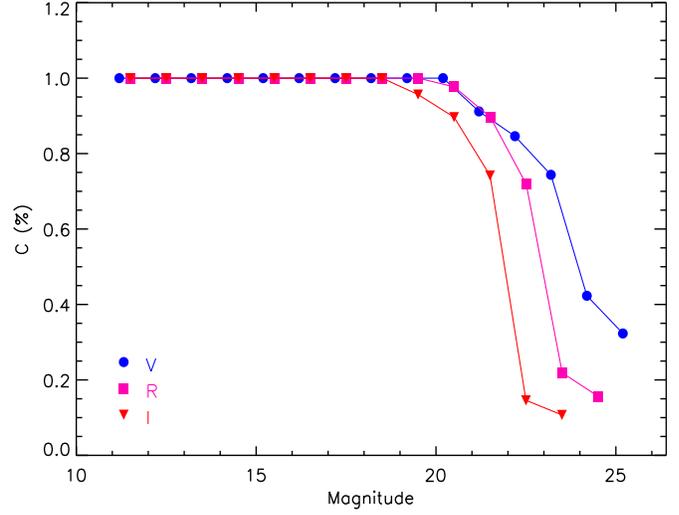}
\caption{Completeness (C) plot for extraction of artificial stars from our ``deep'' 
mosaics (see Sect.~\ref{catalog}) for the three optical bands used in this work.}
\label{fig:compl}
\end{figure}

The completeness of our ``deep'' catalog was derived empirically. 
We used the IRAF \emph{addstar} package to add test stars into the frames and 
then the same PSF-fitting procedure as for the real objects (Sect.~\ref{catalog}) 
to determine what fraction of these artificial stars were recoverable 
as a function of magnitude. This fraction provides a measure of the completeness. 
We inserted 500 artificial sources in each of the three deep frames in the $VRI$ filters; 
this number is statistically significant for our purpose and, at the same time, 
does not alter the crowding in the images. 
The profile for the artificial sources was generated 
by using the same PSF model as used for the source extraction; the positions of the
artificial objects were randomly distributed over the entire area of the mosaic, 
and their magnitudes range uniformly between the detection and the saturation 
limits in each band (Table~\ref{tab:mag_limits}). 
Fig.~\ref{fig:compl} shows the fraction of recovered artificial objects, 
i.e. the completeness (C), as a function of magnitude for each filter. 
We estimate a 100\% completeness level down to $V \approx$20, $R \approx$19.5 and $I \approx$18.5; 
however, objects down to our detection limits at 3$\sigma$ level (Table~\ref{catalog}) 
are recovered with a completeness level better than $\sim$80\%. 
Using the theoretical isochrones ed evolutionary tracks by \citet{Bar98} and 
\citet{Cha00} for the WFI-Cousins system \citep[see Sect.~3.1 by][]{Spe07} 
and considering the age and distance of IC~2391, 
we estimate that the 100\% and 80\% completeness levels 
correspond to objects with M$\approx$0.05~M$_\odot$ and M$\approx$0.03~M$_\odot$, respectively.
Interstellar extinction effects are expected to be negligible (see Sect.~\ref{intro}).

\subsection{Spitzer observations and data reduction \label{IR_data}}

IRAC \citep{Faz04} observations of IC~2391 were obtained on 30 December 2006. 
The 12s high-dynamic-range mode was used to obtain two frames in each position, one with 0.4~sec 
exposure time and one with 10.4~sec. The observation of each field was repeated three 
times with a small offset, providing 31.2~sec integration time for each position. 
The frames were processed using the SSC IRAC Pipeline v15.05, and mosaics were 
created from the basic calibrated data (BCD) frames using a custom IDL program 
(see \citet{Gut08} for details). 
Aperture photometry on these images was carried out using PhotVis version 1.10, 
which is an IDL-GUI based photometry visualization tool. See \citet{Gut04} 
for further details on PhotVis. The radii of the source aperture, and of the 
inner and outer boundaries of the sky annulus were 2.4, 2.4 and 7.2 arcsec, 
respectively. The calibration used large aperture measurements of standard stars. 
The zero point magnitudes of the calibration were 19.6642, 18.9276, 16.8468, and 
17.3909 for channel 1, 2, 3, and 4, respectively. Aperture corrections of 0.21, 
0.23, 0.35 and 0.5~mag were applied for channels 1, 2, 3, and 4 to account for 
the differences between the aperture sizes used for the standard stars and for 
the IC~2391 photometry. The sensitivity of the IRAC observations is sufficient for the 
detection of objects with $K_S\lesssim$14.5, which corresponds to 
the hydrogen-burning limit for cluster members.

MIPS observations in IC~2391 and the relative data reduction and calibration 
have been already described by \citet{Sie07}. 

The standard fluxes at each Spitzer pass-band were derived from the observed magnitudes using 
the absolute flux calibration constants by \citet{Rea05} and \citet{Eng07} 
for IRAC and MIPS/24$\mu$m bands, respectively.

\section{Selection of new cluster members \label{selection}}

The extraction of member candidates in open clusters generally
consists in selecting objects whose position in 
color-magnitude diagrams (CMDs) is located above the
ZAMS shifted to the distance of the cluster 
\citep[see, e.g.,][]{Leg92,Bar01a,Lod05}. 
The selection of IC~2391 member candidates was 
carried out using a similar criterion, refined by exploiting 
estimates of the cluster proper motion and 
the photometry of previously known cluster 
members available from the literature. 
Our selection procedure includes the 
three steps which are now described.

\subsection{Determination of stellar parameters \label{par}}

We first assumed that all the point-like objects detected 
in our images above 3$\sigma$ level are all at the distance of IC~2391 
and simultaneously estimated  
their visual extinction (A$_V$), effective temperature (T$_{\rm eff}$) 
and stellar radius (R$_\star$) by fitting a grid of stellar 
photosphere models to their observed spectral energy distribution (SED). 
The observed SEDs were derived by merging the $VRI$ WFI-Cousins photometry with 
the 2MASS $JHK_S$ photometry; the standard fluxes 
at each pass-band were derived from the observed magnitudes using 
the absolute flux calibration constants by \citet{Joh65} and \citet{Cou76} 
for the optical bands and \citet{Cut03} for the 2MASS bands, respectively. 
The grid of reference SEDs was constructed by using the NextGen and Stardusty 
model spectra by \citet{Hau99} and \citet{All00} 
and the WFI and 2MASS filter transmission curves, 
as prescribed by \citet{Spe07}. Our grid spans the 
T$_{\rm eff}$ range 1700-10000~K and assumes a surface
gravity $\log g = 4.5$, as expected for low-mass objects at the age of IC~2391. 
The reference SEDs were scaled to the distance to IC~2391 (i.e. 150~pc) and 
fitted to the observed SED of each object following 
the prescription by \citet{Spe07} and adopting the extinction law by 
\citet{Car89}\footnote{We also checked that using other extinction laws, 
such as those by \citet{Sav79} and \citet{Wei01}, does not affect 
significantly our results. We adopt the \citet{Car89} extinction law because 
its analytic form is more functional for our purposes.}.  
The A$_V$, T$_{\rm eff}$ and R$_\star$ values of the reference SED minimizing 
the $\chi^2$ of the fit correspond to the best approximation 
of the actual stellar parameters. The stellar luminosity (L$_\star$) 
is derived as $L_\star = 4 \pi \sigma R_{\star 2} T_{eff}^4$, 
where $\sigma$ is the Stefan-Boltzmann constant. 
For more details on the SED fitting procedure we defer the reader to \citet{Spe07}. 
The goodness of our fits can be appreciated from the examples shown in Fig.~\ref{fig_seds}.

As demonstrated by \citet{Spe07}, this parametrization method turns out to
be accurate within about 250\,K in T$_{\rm eff}$ and 1.5\,mag in A$_V$ for objects 
with spectral type later than K5 (i.e T$_{eff} \lesssim$4500~K). 
Such uncertainties translate into an average uncertainty 
of 0.15~dex in the logarithm of luminosity.
This accuracy in the derived physical parameters 
is sufficient for candidate selection purposes. Moreover, 
on the basis of the saturation limits of our survey (Table~\ref{tab:mag_limits}), 
we expect to recover only the low-mass cluster members ($R_C \gtrsim$10~mag, 
i.e. M$\lesssim$1~M$_\odot$ at the age and distance of IC~2391),
for which the parametrization method by \citet{Spe07} can be confidently applied. 
Note also that possible IR excesses at 2MASS
wavelengths due to the presence of hot circumstellar dust does not
appear to be a major issue in our fits. Indeed, the age of IC~2391 is 
consistent with both theoretical and observational timescales of 
disk dissipation and terrestrial planet formation \citep{Cha01,Kle02,Sie07}; 
thus, following the classification by \citet{Lad87} as extended by \citet{Gre94}, 
the cluster population is expected to consist mainly 
of class III objects with no prominent IR excess at near-IR wavelengths. 
Moreover, the same parametrization procedure was applied to 40 confirmed 
members of IC~2391 whose spectral type (later than K5) and 
optical/near-IR photometry were recovered from the literature 
\citep{Pat96,Rol97,Sim98,Pat99,Bar99,Bar01b,Bar04,Cut03}. 
Our fitting procedure yields for these objects T$_{\rm eff}$ and A$_V$ values 
consistent with those derived from spectroscopy within $\sim$250~K and
$\sim$0.7~mag, respectively. This further supports the reliability 
of our parametrization method.

\subsection{Selection of cluster member candidate from the HR diagram \label{sel_HR}}

\begin{figure} 
\centering
\includegraphics[width=9cm,height=6.5cm]{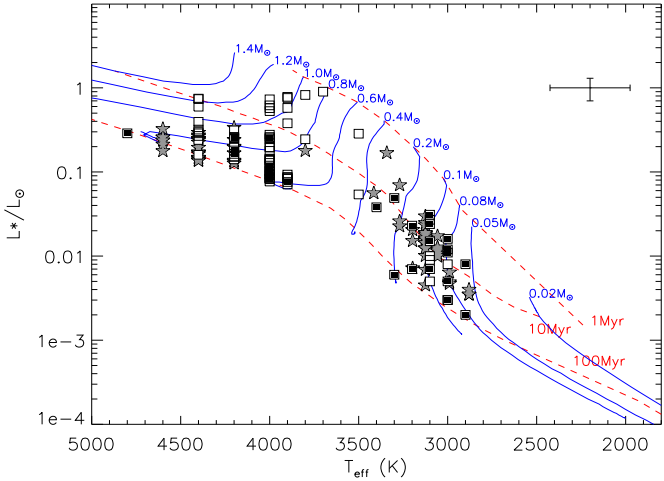}
\caption{HR diagram for 40 low-mass confirmed members of IC~2391 (stars), 
the sample of optical sources whose isochronal age is $<$100~Myr 
(open squares; see Sect.~\ref{sel_HR}) and the final sample 
of 29 member candidates selected in this work (filled squares). 
The continuous and dashed lines are the evolutionary tracks 
and isochrone by \citet{Bar98} and \citet{Cha00}. 
The average error bars on effective temperature and 
luminosity are shown in the top-right.}
\label{HRdiag}
\end{figure}

Having a reliable estimate of T$_{\rm eff}$ and L$_\star$ 
for all the sources in our catalog with T$_{\rm eff} \lesssim$4500~K 
(i.e. 2395 sources), we constructed the Hertzsprung-Russell (HR) diagram 
for these sources and for the 40 confirmed 
members of IC~2391 mentioned above (Sect.~\ref{par}). 
Mass and age of these objects can be now determined  by comparison of 
their location on the HR diagram with a suitable set of evolutionary tracks, 
which in our case are those by \citet{Bar98} and \citet{Cha00}. 
We find that the isochronal age of the 40 low-mass 
cluster members ranges between 10 and 80~Myr, 
with a mean value of 20-30~Myr (Fig.~\ref{HRdiag}); 
this isochronal age is in fair agreement with the 
cluster age estimated on the basis of the lithium depletion technique 
and the main-sequence isochrone fitting 
\citep[30-50~Myr;][]{Mer81,Bas96,Sta98,Bar99,Bar04}. 
However, we observe an age spread between a few Myr and 100~Myr 
(see also Table~\ref{tab_par}) which is larger than that reported in the literature. 
This larger age spread mainly arises from our indetermination on effective temperature and luminosity 
(Sect.~\ref{par}); indeed, stellar parameters inferred from photometry alone are far more 
inaccurate than those derived from spectroscopy, though 
still useful for candidate selection purposes.

New low-mass cluster members are expected to be found in the 
same locus of the HR diagram. Thus, we performed a first-level 
selection of member candidates by picking up those objects 
whose isochronal age is $<$100~Myr; this selection yielded 90 candidates. 
Clearly, we also find a large number 
of source in our catalog for which we cannot assign an age
because their location on the HR diagram is inconsistent with 
the pre-main sequence (PMS) locus when the distance of IC~2391 is assumed; 
these sources are field objects unrelated to the cluster. 

The advantage of using this approach, as compared to frequently
used membership criteria based on CMDs, lies in the simultaneous use of 
the photometric information from the $V$ to the $K_S$ bands;
indeed, this wavelength range samples the peak of the photospheric 
emission of cool, low-mass objects, and enables a highly reliable 
determination of the stellar parameters (Comer\'on et al., in preparation).

\begin{table*}
\caption[ ]{\label{tab_phot} 
Optical and near-IR photometry for the IC~2391 members and candidates of IC~2391 discussed in this work.
The objects with identification number (ID) from 1 to 29 are the new candidates selected in this work (Sect.~\ref{selection}); 
three of these candidates show IR excess emission at 24$\mu$m and are marked by an asterisk. 
The objects with ID from 30 to 37 are the additional 8 cluster members and candidates showing IR 
excess emission discussed in Sect.~\ref{sec_debris}.}
\begin{center}
\tiny
\begin{tabular}{cllcccccccccc}  
\hline
ID    & Designation &  Status$^\wedge$ & Ref.$^{\bullet}$ & R.A.$_{J2000} ^{\dag}$ & Dec.$_{J2000} ^{\dag}$ & $B^{\diamond}$ & $V^{\dag}$ & $R_{\rm c} ^{\dag}$ & $I_{\rm c} ^{\dag}$ & $J^{\ddag}$ & $H^{\ddag}$ & $K_S ^{\ddag}$\\
      &             &   	       &		  & (hh:mm:ss)  	   &  (dd:mm:ss)	    &	  &	&  &  &  &  & \\
\noalign{\medskip}
\hline
\noalign{\medskip}
1  & WFI~J08403964-5250429*	    & HR-C      &    & 08:40:39.643 & $-$52:50:42.86 &  --   & 21.24$\pm$0.09  & 19.82$\pm$0.05 & 17.83$\pm$0.07 & 15.54$\pm$0.05 & 14.86$\pm$0.08 & 14.52$\pm$0.09 \\
2  & WFI~J08404512-5248521 	    & HR-C,PM-C &    & 08:40:45.118 & $-$52:48:52.09 & 17.10 & 16.60$\pm$0.07  & 15.39$\pm$0.03 & 13.97$\pm$0.04 & 12.23$\pm$0.02 & 11.65$\pm$0.02 & 11.42$\pm$0.02 \\
3  & WFI~J08404578-5301052 	    & HR-C      &    & 08:40:45.780 & $-$53:01:05.23 &  --   & 14.50$\pm$0.07  & 13.36$\pm$0.03 & 12.63$\pm$0.04 & 11.24$\pm$0.03 & 10.43$\pm$0.04 & 10.19$\pm$0.05 \\
4  & WFI~J08405518-5257227	    & HR-C      &    & 08:40:55.176 & $-$52:57:22.68 & 15.52 & 14.59$\pm$0.07  & 13.39$\pm$0.03 & 12.58$\pm$0.04 & 10.94$\pm$0.03 & 10.10$\pm$0.03 &  9.84$\pm$0.02 \\
5  & WFI~J08410806-5300053$^{OD}$   & HR-C,X    & a  & 08:41:08.057 & $-$53:00:05.26 &  --   & 18.02$\pm$0.08  & 16.82$\pm$0.06 & 15.27$\pm$0.06 & 13.28$\pm$0.03 & 12.67$\pm$0.06 & 12.45$\pm$0.05 \\					
6  & WFI~J08410814-5255027 	    & HR-C,PM-C &    & 08:41:08.141 & $-$52:55:02.71 & 14.30 & 14.69$\pm$0.07  & 13.45$\pm$0.03 & 12.64$\pm$0.04 & 10.98$\pm$0.04 & 10.08$\pm$0.04 &  9.81$\pm$0.03 \\
7  & WFI~J08410886-5248387*	    & HR-C,X    &    & 08:41:08.856 & $-$52:48:38.74 & 18.77 & 18.74$\pm$0.07  & 17.46$\pm$0.04 & 15.66$\pm$0.06 & 13.52$\pm$0.03 & 12.90$\pm$0.02 & 12.65$\pm$0.03 \\
8  & WFI~J08410937-5302136 	    & HR-C,X    &    & 08:41:09.374 & $-$53:02:13.60 & 17.51 & 17.58$\pm$0.07  & 16.38$\pm$0.04 & 14.74$\pm$0.05 & 12.73$\pm$0.03 & 12.11$\pm$0.03 & 11.91$\pm$0.03 \\
9  & WFI~J08411865-5258549$^{OD}$   & HR-C,X    & a  & 08:41:18.650 & $-$52:58:54.88 & 19.61 & 18.83$\pm$0.07  & 17.64$\pm$0.03 & 16.12$\pm$0.05 & 14.14$\pm$0.03 & 13.58$\pm$0.03 & 13.23$\pm$0.04 \\
10 & WFI~J08412229-5304471 	    & HR-C,X    &    & 08:41:22.291 & $-$53:04:47.14 & 17.77 & 17.16$\pm$0.07  & 15.98$\pm$0.03 & 14.43$\pm$0.05 & 12.47$\pm$0.02 & 11.85$\pm$0.02 & 11.58$\pm$0.02 \\
11 & WFI~J08413933-5252565 	    & HR-C      &    & 08:41:39.326 & $-$52:52:56.53 & 15.76 & 14.93$\pm$0.07  & 13.81$\pm$0.03 & 13.09$\pm$0.04 & 11.60$\pm$0.02 & 10.85$\pm$0.03 & 10.65$\pm$0.02 \\
12 & WFI~J08415360-5257558 	    & HR-C,X    &    & 08:41:53.602 & $-$52:57:55.80 & 18.12 & 17.44$\pm$0.07  & 16.24$\pm$0.03 & 14.72$\pm$0.05 & 12.79$\pm$0.03 & 12.17$\pm$0.02 & 11.93$\pm$0.03 \\
13 & WFI~J08422460-5259575 	    & HR-C,PM-C &    & 08:42:24.605 & $-$52:59:57.48 & 13.71 & 12.71$\pm$0.05  & 11.91$\pm$0.04 & 11.48$\pm$0.05 & 10.54$\pm$0.03 & 10.05$\pm$0.03 &  9.92$\pm$0.02 \\
14 & WFI~J08423071-5257348 	    & HR-C,PM-C,X&   & 08:42:30.715 & $-$52:57:34.78 & 14.95 & 14.63$\pm$0.07  & 13.49$\pm$0.03 & 12.73$\pm$0.04 & 11.39$\pm$0.02 & 10.71$\pm$0.02 & 10.52$\pm$0.02 \\
15 & WFI~J08424731-5309235 	    & HR-C,X    &    & 08:42:47.314 & $-$53:09:23.54 & 17.80 & 20.36$\pm$0.08  & 18.91$\pm$0.05 & 16.79$\pm$0.07 & 14.20$\pm$0.05 & 13.64$\pm$0.05 & 13.24$\pm$0.04 \\
16 & WFI~J08424850-5303187 	    & HR-C,X    &    & 08:42:48.502 & $-$53:03:18.72 & 19.07 & 18.55$\pm$0.07  & 17.26$\pm$0.04 & 15.63$\pm$0.05 & 13.59$\pm$0.02 & 12.99$\pm$0.02 & 12.68$\pm$0.03 \\
17 & WFI~J08425873-5305546 	    & HR-C,PM-C,X&   & 08:42:58.733 & $-$53:05:54.64 & 17.56 & 16.55$\pm$0.07  & 15.33$\pm$0.03 & 13.87$\pm$0.04 & 12.01$\pm$0.02 & 11.38$\pm$0.02 & 11.13$\pm$0.02 \\
18 & WFI~J08425988-5305074 	    & HR-C      &    & 08:42:59.880 & $-$53:05:07.44 &  --   & 20.65$\pm$0.08  & 19.19$\pm$0.05 & 17.17$\pm$0.07 & 14.77$\pm$0.04 & 14.16$\pm$0.04 & 13.81$\pm$0.06 \\
19 & WFI~J08430294-5302259 	    & HR-C      &    & 08:43:02.938 & $-$53:02:25.91 & 15.07 & 13.97$\pm$0.07  & 12.87$\pm$0.03 & 12.10$\pm$0.04 & 10.60$\pm$0.02 &  9.79$\pm$0.02 &  9.59$\pm$0.02 \\
20 & WFI~J08431228-5251411	    & HR-C      &    & 08:43:12.283 & $-$52:51:41.11 &  --   & 22.85$\pm$0.10  & 21.32$\pm$0.07 & 18.92$\pm$0.09 & 16.18$\pm$0.09 & 15.53$\pm$0.13 & 15.45$^{UL}$   \\
21 & WFI~J08431860-5306241 	    & HR-C,PM-C &    & 08:43:18.602 & $-$53:06:24.12 & 13.97 & 13.41$\pm$0.06  & 12.43$\pm$0.03 & 11.80$\pm$0.04 & 10.52$\pm$0.02 &  9.83$\pm$0.02 &  9.67$\pm$0.02 \\
22 & WFI~J08433366-5259549 	    & HR-C      &    & 08:43:33.658 & $-$52:59:54.89 &  --   & 22.71$\pm$0.12  & 20.88$\pm$0.07 & 18.55$\pm$0.08 & 15.80$\pm$0.08 & 15.05$\pm$0.09 & 14.75$\pm$0.11 \\
23 & WFI~J08433843-5246379 	    & HR-C      &    & 08:43:38.431 & $-$52:46:37.92 & 18.09 & 16.96$\pm$0.08  & 15.49$\pm$0.03 & 14.36$\pm$0.04 & 12.38$\pm$0.02 & 11.36$\pm$0.02 & 11.06$\pm$0.02 \\
24 & WFI~J08433939-5307563$^{OD}$   & HR-C      & b  & 08:43:39.389 & $-$53:07:56.32 & 18.38 & 18.08$\pm$0.07  & 16.87$\pm$0.04 & 15.20$\pm$0.05 & 13.14$\pm$0.02 & 12.54$\pm$0.03 & 12.28$\pm$0.02 \\
25 & WFI~J08433945-5248576 	    & HR-C      &    & 08:43:39.451 & $-$52:48:57.56 & 19.74 & 18.79$\pm$0.07  & 17.60$\pm$0.03 & 16.16$\pm$0.04 & 14.38$\pm$0.03 & 13.76$\pm$0.03 & 13.46$\pm$0.05 \\
26 & WFI~J08434640-5257036 	    & HR-C,PM-C &    & 08:43:46.399 & $-$52:57:03.56 & 14.48 & 13.73$\pm$0.06  & 12.77$\pm$0.03 & 12.12$\pm$0.04 & 10.86$\pm$0.03 & 10.23$^{UL}$	& 10.01$^{UL}$   \\
27 & WFI~J08434751-5250136 	    & HR-C,PM-C &    & 08:43:47.508 & $-$52:50:13.63 & 16.01 & 15.92$\pm$0.07  & 14.59$\pm$0.03 & 13.64$\pm$0.03 & 11.88$\pm$0.02 & 10.91$\pm$0.02 & 10.65$\pm$0.02 \\
28 & WFI~J08434909-5307269 	    & HR-C      &    & 08:43:49.090 & $-$53:07:26.94 & 19.51 & 18.64$\pm$0.07  & 17.48$\pm$0.03 & 16.00$\pm$0.04 & 14.13$\pm$0.03 & 13.55$\pm$0.03 & 13.28$\pm$0.04 \\
29 & WFI~J08440199-5310356 	    & HR-C,PM-C &    & 08:44:01.992 & $-$53:10:35.62 & 15.20 & 14.95$\pm$0.06  & 13.94$\pm$0.03 & 13.24$\pm$0.04 & 11.65$\pm$0.04 & 11.09$\pm$0.05 & 10.69$\pm$0.03 \\
\hline
30 & WFI~J08404683-5302487*$^{OD}$  & M      & c  & 08:40:46.826 & $-$53:02:48.73  &  --   & 19.25$\pm$0.06 & 18.36$\pm$0.03 & 17.80$\pm$0.04 & 16.53$\pm$0.21 & 15.81$\pm$0.28 & 15.95$^{UL}$     \\
31 & WFI~J08410019-5249088*         & IR-C   &    & 08:41:00.190 & $-$52:49:08.76  & 17.53 & 16.71$\pm$0.07 & 15.60$\pm$0.03 & 14.92$\pm$0.04 & 13.51$\pm$0.03 & 12.77$\pm$0.03 & 12.56$\pm$0.03  \\
32 & WFI~J08412176-5247289*$^{OD}$  & M      & c  & 08:41:21.756 & $-$52:47:28.90  & 20.10 & 19.33$\pm$0.06 & 18.24$\pm$0.03 & 17.65$\pm$0.04 & 16.43$\pm$0.11 & 15.60$\pm$0.14 & 15.25$^{UL}$     \\
33 & WFI~J08412550-5300154*         & IR-C   &    & 08:41:25.502 & $-$53:00:15.44  &  --   & 18.73$\pm$0.05 & 17.92$\pm$0.04 & 17.49$\pm$0.05 & 16.35$\pm$0.11 &  --	     & 15.60$^{UL}$	\\
34 & WFI~J08413601-5309274*$^{OD}$  & M      &d,e & 08:41:36.010 & $-$53:09:27.40  & 19.37 & 18.52$\pm$0.07 & 17.27$\pm$0.04 & 15.56$\pm$0.05 & 13.43$\pm$0.02 & 12.89$\pm$0.03 & 12.61$\pm$0.03  \\
35 & WFI~J08413888-5259492*         & IR-C   &    & 08:41:38.885 & $-$52:59:49.24  & 17.22 & 17.20$\pm$0.06 & 16.30$\pm$0.03 & 15.73$\pm$0.04 & 14.56$\pm$0.02 & 14.16$\pm$0.03 & 13.95$\pm$0.06  \\
36 & WFI~J08414904-5255192*         & IR-C   &    & 08:41:49.039 & $-$52:55:19.20  & 19.56 & 18.52$\pm$0.06 & 17.46$\pm$0.03 & 16.77$\pm$0.04 & 15.19$\pm$0.05 & 14.61$\pm$0.06 & 14.47$\pm$0.10  \\
37 & WFI~J08415780-5252140*$^{OD}$  & M      & f  & 08:41:57.799 & $-$52:52:13.98  & 14.74 & 13.76$\pm$0.06 & 12.74$\pm$0.03 & 12.16$\pm$0.04 & 11.09$\pm$0.02 & 10.40$\pm$0.02 & 10.27$\pm$0.02  \\
\noalign{\medskip}   						        
\hline               						        
\end{tabular}
\end{center}
$^\wedge$ \footnotesize{M = confirmed cluster member; HR-C = member candidate according to the location on the HR diagram; 
PM-C = proper motion member candidate; X = X-ray sources \citep{Mar05}.; 
IR-C: young stellar object candidate according to the IR properties (Sect.~\ref{sel_debris}).}\\
$^\bullet$ \footnotesize{a) \citet{Rol97}; b) \citet{Pat99}; c) \citet{Dod04}; e) \citet{Mar05}; f) \citet{Sie07} and references therein.}\\
$^\dag$ \footnotesize{From the WFI survey presented in this work.}\\
$^\diamond$ \footnotesize{From the NOMAD catalog \citep{Zac05}. 
Photometric uncertainties are of the order of 0.2-0.3~mag.}\\
$^\ddag$ \footnotesize{From the 2MASS catalog \citep{Cut03}.}\\
* \footnotesize{Object showing IR excess emission at 24$\mu$m.}\\
$^{UL}$ \footnotesize{Flux upper limit.}\\
$^{OD}$ \footnotesize{Other Designation. ID~5: RB729, ID~9: RB955, 2MASSJ08411862-5258546. ID~24: PP9, 2MASSJ08433935-5307556. 
ID~30: D97, 2MASS~J08404679-5302481. ID~32: D112, 2MASSJ08412176-5247291. ID~34: CTIO100, 2MASSJ08413595-5309268. 
ID~37: V368Vel, SMY29, PMM4636, GSC08569-01100, VXRPSPC41, SHJM9, VXRHRI19, \#29.}\\

\end{table*}

\begin{figure} 
\resizebox{\hsize}{!}{\includegraphics[width=2.5cm,height=2cm]{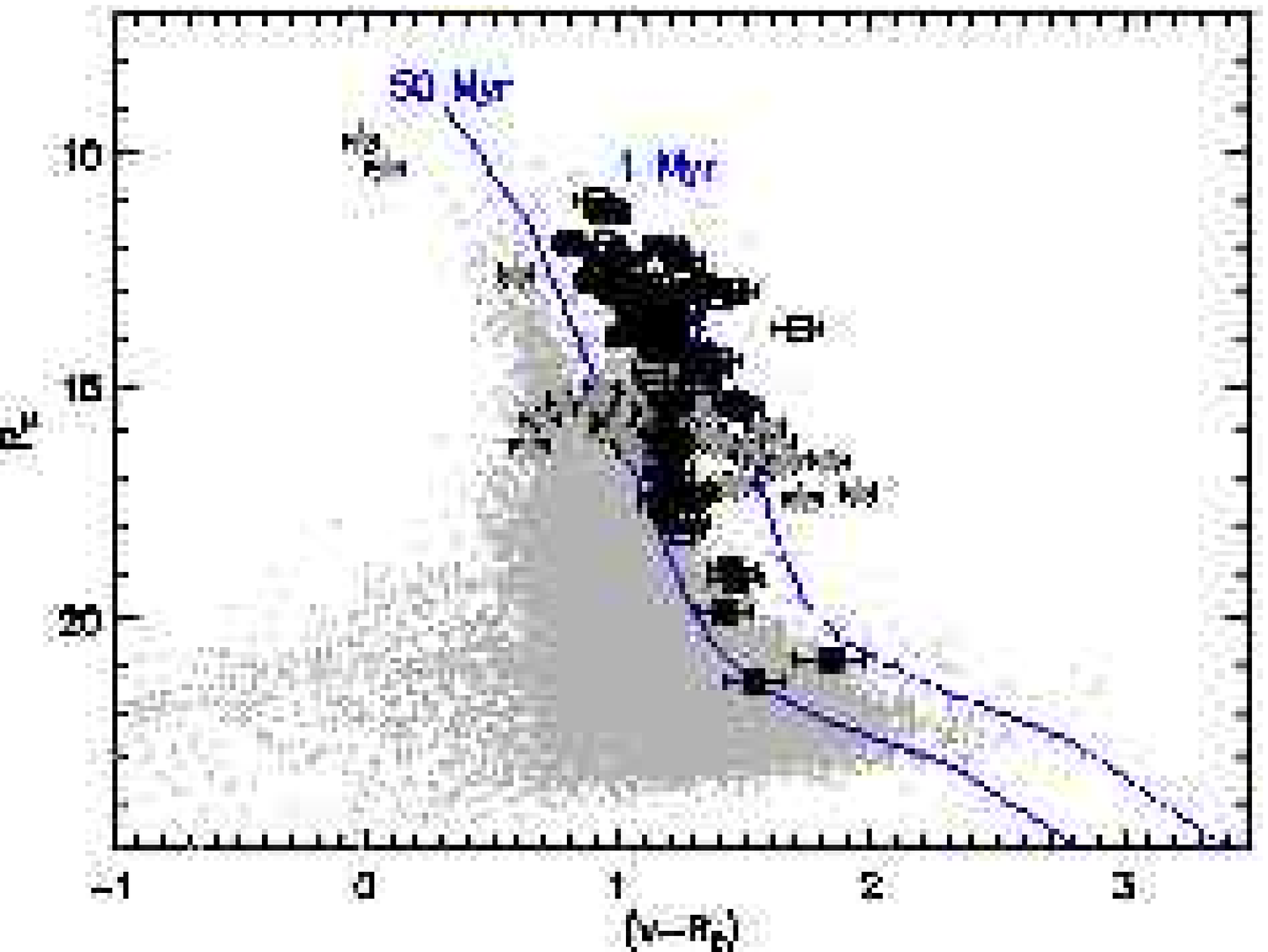}}
\resizebox{\hsize}{!}{\includegraphics[width=2.5cm,height=2cm]{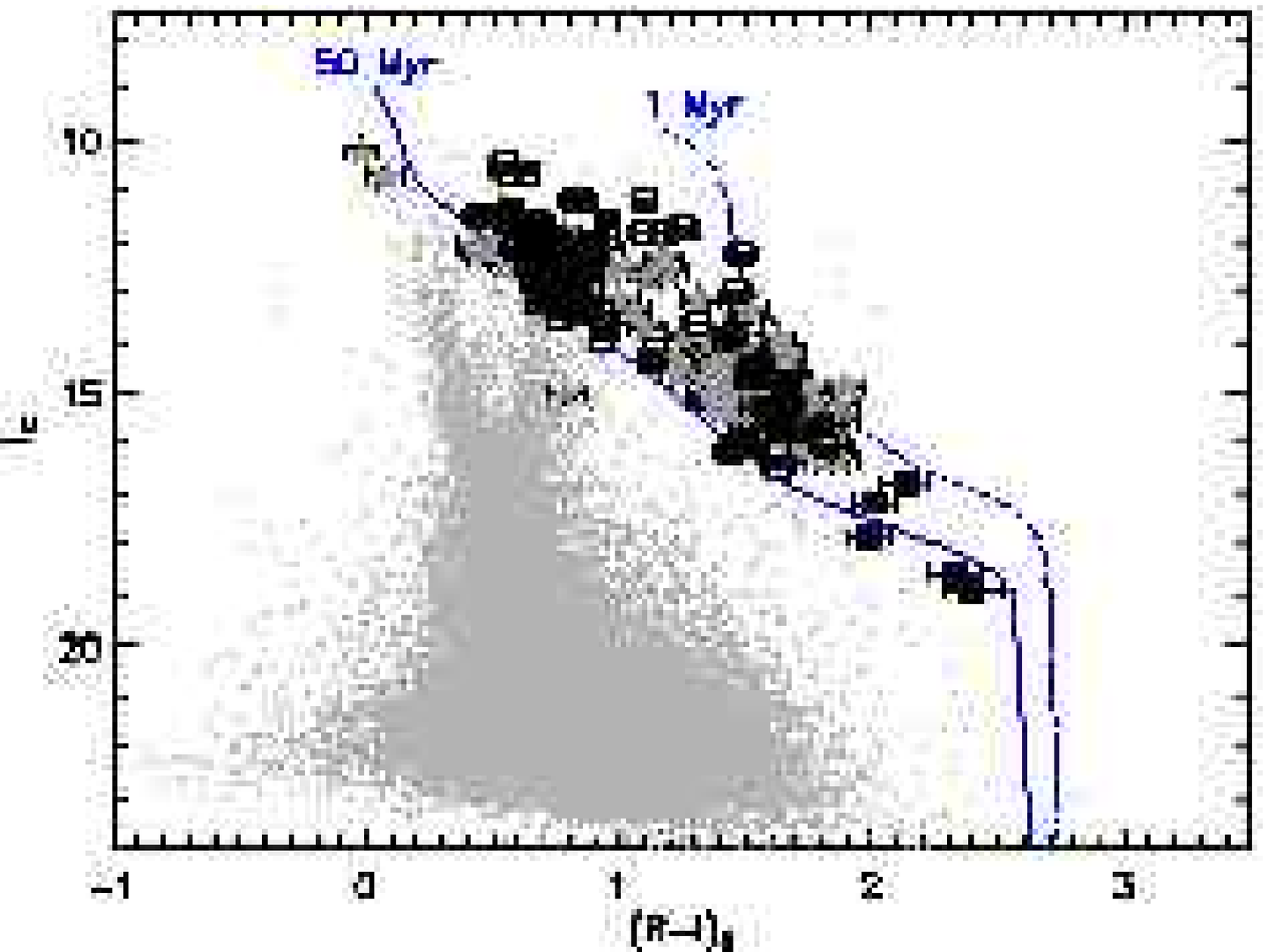}}
\caption{$R_C$ vs. ($V-R_C$) and $I_C$ vs. ($R-I$)$_C$ CMDs for the 
point-like sources in our WFI survey (small gray dots). 
The lines represent the 1 and 50~Myr theoretical isochrones derived 
as explained in Sect.~\ref{sel_pm}, shifted to the 
distance modulus of IC~2391 \citep[5.88 mag;][]{For01}. 
The open squares represent the sample of optical sources whose 
isochronal age is $<$100~Myr (Sect.~\ref{sel_HR}), 
while the filled squares mark the final sample 
of 29 member candidates (Sect.~\ref{sel_pm}). 
The stars are the confirmed cluster members, whose 
photometry was retrieved from the literature \citep{Pat96,Rol97,Sim98,Pat99,Bar01b}. 
Error bars are also drawn; where not visible, they are smaller than the symbol size.}
\label{CMDs}
\end{figure}

\subsection{Refining the sample using proper motion measurements \label{sel_pm}}

Fig.~\ref{CMDs} show the $R_C$ vs. $(V-R_C)$ and $I_C$ vs. $(R-I)_{C}$ 
CMDs for all the point-like optical sources detected 
in IC~2391 above 3$\sigma$ level and the 90 candidates selected 
from the inspection of the HR diagram. The lines in these plots 
are the 1 and 50~Myr theoretical NextGen isochrones by \citet{Bar98} for effective 
temperatures higher than 2500~K and the DUSTY isochrones by 
\citet{Cha00} for lower temperatures; 
these isochrones were transformed into the 
WFI-Cousins system, as described by \citet{Spe07}, and scaled 
to the IC~2391 distance of 150~pc. 
No correction was made for the mean reddening towards IC~2391, 
which is estimated close to zero \citep[E(B-V)=0.006;][]{Pat96}. 
Both diagrams show a wide sequence of candidates with $12 \lesssim R_C \lesssim$15~mag 
and 11$\lesssim I_C \lesssim$14~mag, which is likely due
to the merging of the cluster and field star sequences.  
Indeed, since the galactic latitude of IC~2391 is about $-7^o$, 
the sample of photometrically selected candidates 
is inevitably contaminated by field stars. 

In order to weed out stellar contaminants, proper motions of cluster member 
candidates are generally computed and compared to those of field 
objects \citep[see, e.g.,][]{Cas07}. 
We applied this proper motion analysis to the candidates in our sample 
using two proper motion catalogs: the all-sky Naval Observatory Merged Astrometric Dataset \citep[NOMAD][]{Zac05}, 
reporting proper motions for stars down to $R_C \approx$16~mag with an average 
accuracy of 6~mas/yr, and the catalog recently presented by \citet{Pla07}, 
who obtained proper motions for 6991 stars down to V$\approx$13-16 over a 
$\sim$9~deg$^2$ area around IC~2391 with a precision of about 1.7~mas/yr.
These two catalogs report consistent proper motions for the confirmed cluster members, 
however the NOMAD catalog is photometrically deeper and hence, 
we found a much larger number of matches with our candidate list with respect to the \citet{Pla07} catalog 
Thus, we used the proper motions from NOMAD, where 
70 candidates out of 90 in our original list (Sect.~\ref{sel_HR}) have proper motion estimates. 
Because of the low precision, we cannot calculate a probability
of membership by the standard method as described in \citet{Mor03}. 
Consequently, we only selected objects to be proper
motion members if their proper motions fall 
within 3$\sigma$ of the cluster mean proper motion 
\citep[$pm(RA)\approx$-24~mas/yr and $pm(DEC)\approx$+24~mas/yr;][]{Kha01,Dod04}. 
This criterion yielded only 9 proper motion member candidates out of 70.
Adding these objects to the 20 fainter candidates whose proper motion are not available, 
we end up with a sample of 29 new member candidates of IC~2391. 
Of course, we still expect a certain level of contamination, especially among 
the candidates lacking proper motion estimates; this issue is discussed in Sect.~\ref{contamination}.


\subsection{Statement on the selection results \label{sel_results}}

Our selection method allowed us to identify 29 new candidate members of IC~2391: 
9 selected by photometry and astrometry, and 20 selected by photometry, for which no PM data
is available.
In Table~\ref{tab_phot} and Table~\ref{tab_par} we report the optical 
and near-IR photometry for these candidates and their stellar parameters 
as explained in Sect.~\ref{par}; the object designation follows the scheme 
already adopted by \citet{Spe07}. 
A few of these objects coincide with member candidates selected in previous 
photometric surveys \citep{Rol97, Pat99}, as indicated 
in Table~\ref{tab_phot} (column 5). Moreover, 10 of our candidates, i.e. 34\%, 
are X-ray sources recently detected with XMM-EPIC by \citet{Mar05}; 
these 10 objects are marked with an ``X'' in Table~\ref{tab_phot} (column 4). 
Note that the XMM-EPIC pointing in IC~2391 is centered at R.A.$\approx 8^h 42^m 00^s$ and 
Dec.$\approx -53^d 00' 36''$ and the instrument field of view has a diameter of 30~arcmin. 
Thus, the great majority of our candidates falls in the area observed by \citet{Mar05} and 
the reason why the 66\% of them were not detected must be mainly ascribed to their 
faint X-ray emission \citep[i.e. below the XMM-EPIC sensitivity limits;][]{Ehl08}.

In addition, we correctly recovered 32 previously known cluster members.
We note that these 32 objects represent the $\sim$40\% of the IC~2391 known 
population falling in the area observed with WFI; 
the remaining 60\% are either saturated in our 
WFI images or do not have complete $VR_CI_CJHK_S$ 
photometric data-set in our merged catalog, 
so we could not apply our selection criteria. 
However, using their magnitudes and proper motions from the 
the literature \citep{Pat96,Rol97,Sim98,Pat99,Bar99,Bar01b,Bar04,Cut03,Zac05,Pla07}, 
they would fall within our selection (see Fig.~\ref{HRdiag} and Fig.~\ref{CMDs}).

All these arguments support the reliability of our selection method.

\subsection{Multiple visual systems \label{binarity}}

In order to identify possible multiple systems, we have visually inspected the 32 IC~2391 members falling 
in the area observed with WFI and the proposed 29 candidates in all the available images. 
Following the prescription by \citet{Rei93}, we perform this search for visual 
neighbors in the separation range 1.2-5 arcsec.
The lower limit of this range was fixed considering the typical seeing of our images (Sect.~\ref{obs}). 
The upper limit is imposed to avoid contamination from fore- and background stars. 
This cut-off is somewhat arbitrary; we may justify this choice
considering that the typical star-to-star separation 
on our images is $\sim$5 arcsec and, when a visual companion is found within this distance, 
it mostly lies conspicuously close to the primary in an otherwise rather empty space.

22 objects in our sample are found to have neighbors within less than 5 arcsec. 
However, a nearby star is accepted as a possible companion only if it shows indications of youth.
2 of these neighbors have complete $VR_CI_CJHK_S$ photometric data-set but do not fulfill our 
membership criteria (Sect.~\ref{selection}) and hence they were rejected. 
The remaining 20 neighbors are too faint and do not have near-IR detection in 2MASS, 
so the only possible indication of youth comes from their position on the optical CMDs. 
We found that 7 out of these 20 neighbors have optical magnitudes and colors consistent or marginally consistent 
with those of IC~2391 members\footnote{This membership criterion is explained in Sect.~\ref{sel_pm}.}. 
These 7 objects, together with their possible primary 
companion and the relative angular separation, are reported in Table~\ref{tab_bin}.
Note that at the distance of IC~2391 ($\sim$150~pc), the range of angular separation 
covered by these possible visual binaries corresponds to a linear separation of 350-700 AU, 
which is reasonable for physically bound young low-mass stars \citep{Rei93}.
For the member candidate ID~6 two neighbors were found within the 5~arcsec radius; 
both these neighbors have optical magnitudes and colors consistent with membership to the cluster 
and hence this might be a triple system.

\begin{table*}
\caption[ ]{\label{tab_par} 
Estimates of the stellar parameters for the IC~2391 members and candidates of IC~2391 discussed in this work.
The objects with identification number (ID) from 1 to 29 are the new candidates selected in this work (Sect.~\ref{selection}); 
three of these candidates show IR excess emission at 24$\mu$m and are marked by an asterisk. 
The objects with ID from 30 to 37 are the additional 8 cluster members and candidates showing IR 
excess emission discussed in Sect.~\ref{sec_debris}.}
\begin{center}
\scriptsize
\begin{tabular}{cccccccc}  
\hline
ID    & Designation & A$_V$  & T$_{\rm eff} ^{\bullet}$  & L$_\star$  & R$_\star$  & Mass$^{\dag}$  & Age$^{\dag}$ \\
      &             & (mag)  &  (K)                      &(L$_\odot$) &(R$_\odot$) &(M$_\odot$)     & (Myr)        \\
\noalign{\medskip}
\hline
\noalign{\medskip}
1  & WFI~J08403964-5250429*& 1.63$\pm$0.13 & 3000 & 0.003$\pm$0.001 & 0.19$\pm$0.04 & 0.08$\pm$0.02 & 56$\pm$20 \\
2  & WFI~J08404512-5248521 & 0.01$\pm$0.06 & 3400 & 0.038$\pm$0.017 & 0.56$\pm$0.13 & 0.25$\pm$0.05 & 14$\pm$5  \\
3  & WFI~J08404578-5301052 & 0.97$\pm$0.07 & 4200 & 0.164$\pm$0.075 & 0.76$\pm$0.18 & 0.75$\pm$0.10 & 45$\pm$20 \\
4  & WFI~J08405518-5257227 & 1.80$\pm$0.06 & 4200 & 0.256$\pm$0.118 & 0.86$\pm$0.22 & 0.85$\pm$0.10 & 32$\pm$10 \\
5  & WFI~J08410806-5300053 & 0.19$\pm$0.10 & 3100 & 0.015$\pm$0.007 & 0.43$\pm$0.10 & 0.07$\pm$0.02 &  6$\pm$3  \\
6  & WFI~J08410814-5255027 & 1.96$\pm$0.07 & 4200 & 0.261$\pm$0.120 & 0.97$\pm$0.22 & 0.90$\pm$0.10 & 25$\pm$8  \\
7  & WFI~J08410886-5248387*& 0.50$\pm$0.08 & 3000 & 0.012$\pm$0.006 & 0.42$\pm$0.09 & 0.07$\pm$0.02 &  6$\pm$3  \\
8  & WFI~J08410937-5302136 & 0.26$\pm$0.07 & 3100 & 0.024$\pm$0.011 & 0.54$\pm$0.12 & 0.13$\pm$0.06 &  7$\pm$2  \\
9  & WFI~J08411865-5258549 & 0.05$\pm$0.08 & 3100 & 0.007$\pm$0.003 & 0.28$\pm$0.06 & 0.11$\pm$0.04 & 32$\pm$10 \\
10 & WFI~J08412229-5304471 & 0.03$\pm$0.07 & 3100 & 0.031$\pm$0.014 & 0.61$\pm$0.14 & 0.11$\pm$0.04 &  4$\pm$2  \\
11 & WFI~J08413933-5252565 & 0.42$\pm$0.06 & 4000 & 0.093$\pm$0.043 & 0.64$\pm$0.15 & 0.62$\pm$0.08 & 91$\pm$40 \\
12 & WFI~J08415360-5257558 & 0.21$\pm$0.07 & 3200 & 0.023$\pm$0.011 & 0.50$\pm$0.12 & 0.20$\pm$0.04 & 18$\pm$6  \\
13 & WFI~J08422460-5259575 & 0.07$\pm$0.06 & 4800 & 0.288$\pm$0.133 & 0.78$\pm$0.18 & 0.85$\pm$0.10 & 74$\pm$30 \\
14 & WFI~J08423071-5257348 & 0.05$\pm$0.06 & 4000 & 0.102$\pm$0.047 & 0.67$\pm$0.15 & 0.70$\pm$0.10 & 45$\pm$20 \\
15 & WFI~J08424731-5309235 & 1.74$\pm$0.10 & 2900 & 0.008$\pm$0.004 & 0.35$\pm$0.08 & 0.05$\pm$0.02 &  6$\pm$3  \\
16 & WFI~J08424850-5303187 & 0.03$\pm$0.07 & 3000 & 0.011$\pm$0.005 & 0.38$\pm$0.09 & 0.08$\pm$0.02 &  9$\pm$4  \\
17 & WFI~J08425873-5305546 & 0.22$\pm$0.06 & 3300 & 0.049$\pm$0.023 & 0.68$\pm$0.16 & 0.25$\pm$0.05 &  9$\pm$4  \\
18 & WFI~J08425988-5305074 & 1.74$\pm$0.10 & 3000 & 0.005$\pm$0.002 & 0.26$\pm$0.06 & 0.07$\pm$0.02 & 18$\pm$6  \\
19 & WFI~J08430294-5302259 & 0.63$\pm$0.06 & 4000 & 0.252$\pm$0.116 & 1.05$\pm$0.24 & 0.90$\pm$0.10 & 18$\pm$6  \\
20 & WFI~J08431228-5251411 & 3.19$\pm$0.19 & 2900 & 0.002$\pm$0.001 & 0.18$\pm$0.04 & 0.07$\pm$0.02 & 64$\pm$20 \\
21 & WFI~J08431860-5306241 & 0.18$\pm$0.06 & 4200 & 0.253$\pm$0.117 & 0.95$\pm$0.22 & 0.85$\pm$0.10 & 32$\pm$10 \\
22 & WFI~J08433366-5259549 & 3.51$\pm$0.16 & 3000 & 0.003$\pm$0.001 & 0.20$\pm$0.05 & 0.09$\pm$0.02 & 64$\pm$20 \\
23 & WFI~J08433843-5246379 & 2.53$\pm$0.06 & 3900 & 0.076$\pm$0.035 & 0.61$\pm$0.14 & 0.60$\pm$0.08 & 90$\pm$40 \\
24 & WFI~J08433939-5307563 & 0.02$\pm$0.07 & 3000 & 0.016$\pm$0.007 & 0.47$\pm$0.11 & 0.08$\pm$0.02 &  6$\pm$3  \\
25 & WFI~J08433945-5248576 & 0.12$\pm$0.08 & 3300 & 0.006$\pm$0.003 & 0.23$\pm$0.05 & 0.15$\pm$0.04 & 80$\pm$40 \\
26 & WFI~J08434640-5257036 & 0.12$\pm$0.10 & 4200 & 0.180$\pm$0.083 & 0.80$\pm$0.19 & 0.75$\pm$0.08 & 50$\pm$20 \\
27 & WFI~J08434751-5250136 & 2.00$\pm$0.06 & 4000 & 0.110$\pm$0.051 & 0.69$\pm$0.16 & 0.70$\pm$0.08 & 45$\pm$20 \\
28 & WFI~J08434909-5307269 & 0.01$\pm$0.08 & 3200 & 0.007$\pm$0.003 & 0.26$\pm$0.06 & 0.11$\pm$0.03 & 32$\pm$10 \\
29 & WFI~J08440199-5310356 & 0.41$\pm$0.07 & 4000 & 0.083$\pm$0.038 & 0.60$\pm$0.14 & 0.62$\pm$0.08 & 80$\pm$40 \\
\hline
30 & WFI~J08404683-5302487*& 0.16$\pm$0.17 & 4200         & 2.128$\pm$0.001 & 2.77$\pm$0.05 & 1.40$\pm$0.20 &  5$\pm$2	      \\  %
31 & WFI~J08410019-5249088*& 0.77$^\ddag$  & 3800$^\ddag$ & 0.185$^\ddag$   & 0.97$^\ddag$  & 0.75$^\ddag$  &  20$^\ddag$     \\  
32 & WFI~J08412176-5247289*& 0.22$\pm$0.15 & 4300         & 1.087$\pm$0.001 & 1.80$\pm$0.01 & 1.40$\pm$0.20 &  6$\pm$3	      \\  %
33 & WFI~J08412550-5300154*& 0.13$^\ddag$  & 4300$^\ddag$ & 0.700$^\ddag$   & 2.48$^\ddag$  & 1.20$^\ddag$  &  10$^\ddag$      \\  
34 & WFI~J08413601-5309274*& 0.17$\pm$0.07 & 3000         & 0.012$\pm$0.006 & 0.39$\pm$0.09 & 0.07$\pm$0.02 &  7$\pm$3	       \\
35 & WFI~J08413888-5259492*& 0.44$^\ddag$  & 4100$^\ddag$ & 0.101$^\ddag$   & 1.13$^\ddag$  & 0.65$^\ddag$  &  80$^\ddag$      \\  
36 & WFI~J08414904-5255192*& 0.05$^\ddag$  & 3700$^\ddag$ & 0.127$^\ddag$   & 0.95$^\ddag$  & 0.65$^\ddag$  &  16$^\ddag$      \\  
37 & WFI~J08415780-5252140*& 0.24$\pm$0.2  & 4000         & 0.203$\pm$0.069 & 0.74$\pm$0.16 & 0.85$\pm$0.10 &  25$\pm$8	       \\ 
\noalign{\medskip}
\hline
\end{tabular}
\end{center}
$^\bullet$ \footnotesize{The uncertainty on T$_{\rm eff}$ is assumed to be 250~K on average (see Sect.~\ref{par}).}\\
$^\dag$ \footnotesize{From the evolutionary models by \citet{Bar98} and \citet{Cha00}.}\\
$^\ddag$ \footnotesize{From the \citet{Rob06} models for young stellar objects (see Sect.~\ref{note_obj}).}\\
* \footnotesize{Object showing IR excess emission at 24$\mu$m.}\\
\end{table*}

\begin{table*}
\caption[ ]{\label{tab_bin}  Possible multiple systems in IC~2391 (see Sect.~\ref{binarity}).}
\begin{center}
\scriptsize
\begin{tabular}{c|ccc|c|cccccc}  
\hline
  		   &              & Primary star    &                &  Separation & 	   &             & Neighbor star &             &          	\\
		   &    ID        & Designation     &	Ref.$^{\dag}$& (arcsec)    &  R.A. & Dec.	 &    $V$        & $R_{\rm c}$ & $I_{\rm c}$  \\	   
\hline 
Previously known   &   -   & Cl*~IC~2391~CTIO~126   &  	a	     &    3.1	 & 08:42:49.385  & $-$52:52:16.96 &  18.25$\pm$0.07 & 17.13$\pm$0.03 & 16.46$\pm$0.04 \\
cluster members    &   -   & Cl*~IC~2391~CTIO ~25   &  	a	     &    2.8	 & 08:42:46.025  & $-$52:46:52.64 &  20.02$\pm$0.07 & 18.82$\pm$0.03 & 17.88$\pm$0.06 \\
\hline
                   &   6   &  WFI~J08410814-5255027 &  	b	     &    3.9	 & 08:41:07.726  & $-$52:55:03.61 &  16.00$\pm$0.06 & 14.98$\pm$0.03 & 14.32$\pm$0.04 \\ 
New member         &   "   &         "              &  	b	     &    3.5	 & 08:41:08.489  & $-$52:55:04.19 &  15.72$\pm$0.05 & 15.02$\pm$0.04 & 14.70$\pm$0.05 \\ 
candidates         &  26   &  WFI~J08434640-5257036 &  	b	     &    4.5	 & 08:43:46.814  & $-$52:57:06.12 &  16.39$\pm$0.05 & 15.68$\pm$0.04 & 15.26$\pm$0.05 \\ 
                   &  27   &  WFI~J08434751-5250136 &  	b	     &    3.3	 & 08:43:47.844  & $-$52:50:12.48 &  20.76$\pm$0.07 & 19.45$\pm$0.03 & 18.52$\pm$0.04 \\ 
                   &  29   &  WFI~J08440199-5310356 &  	b	     &    2.4	 & 08:44:02.071  & $-$53:10:37.92 &  15.48$\pm$0.06 & 14.47$\pm$0.03 & 13.77$\pm$0.04 \\ 

\noalign{\medskip}   
\hline               
\end{tabular}
\end{center}
$^\dag$ \footnotesize{a) \citet{Bar01b}; b) This work.}\\
\end{table*}

\section{Estimating the contamination \label{contamination}}

In Sect.~\ref{sel_pm} we adopted the proper motion membership criterion 
to weed out stellar contaminants in the magnitude range $R_C \lesssim$16~mag.
We could not perform this analysis for fainter candidates 
because their proper motion are unknown and hence, we still expect 
a certain level of contamination by field stars in our candidates sample. 

To estimate the degree of contamination we have used optical imaging of a 
nearby field available from the ESO archive. This field was observed with WFI in the $VRI$ filters and 
is centered at R.A.=08$^h$:43$^m$:53$^s$ and Dec.=$-$52$^o$:47$^m$:33$^s$,
i.e. about 0.9~deg north-east from the center of IC~2391; 
since the cluster has an apparent diameter of about 50~arcmin \citep{Kha05}, 
we expect this area to be more or less devoid of cluster members and we can use it as a control field.
The exposure time of the control field images is 600~sec in all the filters; 
this implies that all the source brighter than $R \approx$15 are saturated, 
while sources with 15$\lesssim R \lesssim$21 are detected with a sigma-to-noise 
ratio better than 10. Thus, using these data, we can estimated the level of contamination 
at least in the photometric range spanned by our candidates with no proper motion determination. 
By using the same procedure outlined in Sect.~\ref{selection},
we found 12 objects in our control field satisfying our photometric cluster membership criteria. 
Therefore, we estimate that the 17 objects with 15$\lesssim R \lesssim$21 cataloged by us 
as candidate cluster members (Table~\ref{tab_phot}) still have a probability up to $\sim$70\% of being spurious.

In order to further verify the amount of expected contamination 
in all the photometric range covered by our candidates, i.e.  12$\lesssim R_C \lesssim$21, 
we have also used a different approach. 
Nearly all members of IC~2391 are 
located in the region of the HR diagram approximately defined by 
the 10~Myr and 100~Myr isochrones (Fig.~\ref{HRdiag}). 
In particular, all our candidates are confined in the region defined by 
T$_{\rm eff}$=2900-4800~K and L$_\star$=0.002-0.4~L$_\odot$ 
(i.e. absolute visual magnitude M$_V$=12-6~mag); 
thus, in order to estimate the contamination level of our candidate sample, 
we have to obtain an estimate of the numbers of stars
unrelated to the cluster that may be expected to appear 
in this region of the HR diagram. 
The number of interloping stars can be probed 
by using analytic models of the Galactic stellar distribution, i.e. simulations of 
the expected properties of stars seen towards a given direction 
of the Galaxy over a given solid angle. 
We performed this exercise by using the Galaxy model by \citet{Rob03} and their on-line 
tool\footnote{For details on the adopted Galaxy model we defer the reader to \citet{Rob03}. 
The on-line simulation tool is available at: http://bison.obs-besancon.fr/modele/}. 
In the temperature range of our candidates, foreground stars are 
expected to be main sequence cool dwarfs, whereas red giants 
are expected to dominate the background population. 
Assuming a cluster distance of 150-200~pc, we expect 
some 14 foreground dwarfs in the $\sim$30$\times$30 square 
arcmin area observed in IC~2391 with apparent $R_C$ magnitude 
between 12 and 21 and spectral types MK, i.e. the ranges estimated 
for our candidates (see Table~\ref{tab_phot} and \ref{tab_par}). 
No background giants are expected to be found in the locus 
occupied by the cluster members because all such stars appear 
much brighter than IC~2391 members in the same effective temperature 
range. We thus conclude that only foreground cool main sequence stars can
contribute noticeably to the contamination of our candidate sample, 
with the expected contamination level being 48\%.
Note that this value is very close to the pollution rate of 50\% estimated 
by \citet{Bar01b} for the same area in IC~2391.

Another possible source of contamination is represented
by galaxies that may have colors similar to those of PMS objects. 
However, such bright galaxies are \emph{a priori} excluded from our selection, 
because our PSF extraction method allows an efficient identification/removal of extended 
objects (Sect.~\ref{catalog}). 
The only source of contamination may be the point-like extra-galactic objects, i.e. 
QSO's. According to recent QSO's number counts \citep{Ric05,Fon07}, 
the predicted number of QSO's per square degree become significant 
(i.e. greater than $\sim$10/deg$^2$) beyond $V \approx$22~mag, while the majority of 
our candidates are brighter. Thus, extra-galactic objects 
might affect our candidate sample by no more than 3\%.

In conclusion, the expected contamination of our 
candidate sample due to field objects is $\sim$50\%.

%

\section{On the sub-stellar population in IC~2391 \label{IMF_BD}}

The first and only estimate of the IC~2391 mass spectrum in the very low-mass domain has been derived by \citet{Bar04}. 
Independently of the assumed cluster age between 25 and 50~Myr, they found an 
index of the power-law mass function in the mass range 0.072-0.5~M$_\odot$ 
equal to 0.96$\pm$0.12, consistent with the value reported for other 
open clusters \citep[e.g.][]{Bou98,Bar01a,Bar02,Bar03}. 
Below 0.07~M$_\odot$, the cluster mass spectrum presents a sudden drop below 
which the authors partially explain by the lack 
of completeness of their survey beyond $I \approx$18.5 
(i.e. 0.05~M$_\odot$ for cluster members). 

Although our survey is complete at 100\%  and 80\% levels down to 0.05~M$_\odot$ and 
0.03~M$_\odot$, respectively, and covers 60\% of the sky area spanned by the cluster, 
we found only an handful of candidate substellar members.
In numbers, considering the IC~2391 populations updated to January 2008 
\citep[i.e. $\sim$180 objects;][]{Pat96,Rol97,Sim98,Pat99,Bar99,Bar01b,Bar04,Pla07}, 
there are 14 members with T$_{\rm eff} \lesssim$3100\,K, 
i.e. below the approximate substellar limit for 30-50\,Myr old objects. Thus, 
the fraction of sub stellar objects, 
$R_{ss}=\frac{N(0.02 M_{\odot}-0.08 M_{\odot})}{N(0.08 M_{\odot}-10 M_{\odot})}$, 
in this cluster is $\sim$8\%; by adding our 12 BD candidates 
(i.e. candidates with T$_{\rm eff} \leq$3100\,K; 
see Table~\ref{tab_par}), this fraction would increase to $\sim$15\%. 
However, all our BD candidates have $R$-magnitudes in the range 15$\lesssim R \lesssim$21, 
where we expect a high level of contamination (70\%, see Sect.~\ref{contamination}).
Thus, the true value of $R_{ss}$ is expected to be very close to 10\%. 
Moreover, our physical parametrization indicates 
that all our BD candidates have T$_{\rm eff} \gtrsim$2900~K 
and masses very close to the hydrogen-burning limit. 
Thus, the results of our deeper optical survey 
confirm the lack of members with spectral type later than 
M7-M8 in IC~2391 already pointed out by \citet{Bar04}. 
A similar behavior has been observed in the nearly coeval open clusters 
$\alpha$~Per ($\sim$50~Myr) and NGC~2547 ($\sim$25~Myr) 
by \citet{Bar02} and \citet{Jef04}, respectively. 
Thus, as already concluded by \citet{Bar04}, one of the most solid hypothesis 
explaining the BD deficit in IC~2391 and other open clusters 
remains that by \citet{Dob02} and \citet{Jam03}, i.e. the onset 
of larger-size dust grain formation in the upper atmosphere 
of objects with spectral types M7-M8 or later, which would be responsible 
of the local drop in the shape of the luminosity-mass relation. 

Of course, a number of alternative hypothesis might be also invoked.
For instance, simulations of the dynamics of BD populations in open clusters, which 
consider the effects of a large range of primordial binary populations, 
have also shown that the majority of BDs might be contained within primordial
binary systems which then hides a large proportion of them from detection \citep{Ada02}. 
Alternatively, the deficit of BDs in IC~2391 might be due to 
mass segregation/evaporation effects. 
Dynamical mass segregation acts on a time-scale of the order of the relaxation time 
of a cluster \citep{Bon98}, which for clusters of the size of IC~2391 is
of the order of 1~Myr \citep[][ p. 390]{Bin87}. 
Indeed, \citet{Sag89} found indication that the higher mass stars in IC~2391 
have lower velocity dispersions than the lower mass stars.


\section{On the debris disk population in IC~2391 \label{sec_debris}}

The first survey for debris disks in IC~2391 
was recently performed by \citet{Sie07} using MIPS on board of 
Spitzer. The central square degree of IC~2391 was observed at 
24$\mu$m, with the aim of studying the incidence of debris disks in the cluster. 
The authors report IR excess indicative of debris disks around 
8 previously known members of the cluster, namely 1 A star, 
6 FGK stars and 1 M dwarf, and estimate a debris disk fraction of 
10$^{+17} _{-3}$ for BA stars and 31$^{+13} _{-9}$ for FGK stars.

In this section we further investigate the debris disk population 
in IC~2391, using our optical data in combination with 2MASS data, 
the MIPS catalog by \citet{Sie07} and additional IR data from IRAC. 
This dataset is suitable to probe disk radii from $\sim$0.1~AU 
up to $20$~AU depending on the stellar mass (Mer\'in et al. in preparation).
While the work by \citet{Sie07} focuses only on the previously 
known cluster members, we extend the search for objects with IR excess emission 
to all the sources detected at 24$\mu$m having a counterpart in our optical WFI images. 
This allow us to identify new candidate KM-type members of the cluster bearing 
debris disks.
We remind the reader that our search is spatially limited to the overlapping area observed with 
WFI and Spitzer (Fig.~\ref{fig:obs}). 
Moreover, the selected debris disk candidates must be 
spectroscopically investigated to definitively assess their membership 
to the cluster. Indeed, the experience with 
the Spitzer \emph{Core to Disk} Legacy Survey \citep[hereafter c2d;][]{Eva03} 
teaches us that samples of objects showing mid-IR excess in star forming regions 
could be highly contaminated by background post-main sequence stars, 
mostly in the direction of the galactic plane. 
This contamination might account for up to 30\% of those objects and could be 
even higher in the case of IC~2391, which has a galactic latitude (b$\approx$-7$^o$) 
lower than the typical c2d fields. 
For all these reasons, our search is intended to give a deeper 
insight into the candidate debris disk population in IC2391.

\subsection{Selection of debris disk candidates \label{sel_debris}}

\begin{table*}
\caption[ ]{\label{tab_spitzer} IRAC and MIPS/24$\mu$m magnitudes and SED slope ($\alpha_{[K \& 24\mu m]}$) 
for the 10 objects showing IR excess emission at 24$\mu$m identified in this work.}
\begin{center}
\small
\begin{tabular}{ccccccccc}  
\hline
ID  & Designation & Status$^\wedge$ & IRAC/3.6$\mu$m & IRAC/4.5$\mu$m & IRAC/5.8$\mu$m  & IRAC/8.0$\mu$m  & MIPS/24$\mu$m  & $\alpha_{[K_S \& 24\mu m]}$  \\
\noalign{\medskip}
\hline
\noalign{\medskip}
1  & WFI~J08403964-5250429	  & HR-C    & 14.26$\pm$0.01 & 14.02$\pm$0.01 & 14.13$\pm$0.08 & 14.11$\pm$0.13 & 11.07$\pm$0.25 &  -1.63  \\ 
30 & WFI~J08404683-5302487	  & M	    & 15.76$\pm$0.04 & 15.61$\pm$0.06 & 15.92$\pm$0.34 & 15.90$\pm$0.98 & 10.71$\pm$0.18 &  -0.89  \\ 
31 & WFI~J08410019-5249088	  & IR-C    & 12.46$\pm$0.01 & 12.54$\pm$0.01 & 12.47$\pm$0.02 & 12.39$\pm$0.04 & 10.31$\pm$0.14 &  -2.07  \\ 
7  & WFI~J08410886-5248387	  & HR-C,X  & 12.32$\pm$0.01 & 12.21$\pm$0.01 & 12.22$\pm$0.02 & 12.10$\pm$0.03 & 11.42$\pm$0.47 &  -2.45  \\ 
32 & WFI~J08412176-5247289	  & M	    & 15.58$\pm$0.03 & 15.59$\pm$0.05 & 15.26$\pm$0.23 & 16.88$\pm$1.60 & 11.41$\pm$0.32 &  -1.44  \\ 
33 & WFI~J08412550-5300154$^\dag$ & IR-C    & 15.88$\pm$0.05 & 15.93$\pm$0.06 & --    	    & --	     & 10.54$\pm$0.17 &  -0.96  \\ 
34 & WFI~J08413601-5309274	  & M	    & 12.35$\pm$0.01 & 12.24$\pm$0.01 & 12.14$\pm$0.02 & 12.17$\pm$0.03 & 11.44$\pm$0.30 &  -2.46  \\ 
35 & WFI~J08413888-5259492	  & IR-C    & 13.70$\pm$0.01 & 13.58$\pm$0.02 & 13.28$\pm$0.05 & 12.95$\pm$0.06 &  9.86$\pm$0.09 &  -1.34  \\ 
36 & WFI~J08414904-5255192	  & IR-C    & 14.07$\pm$0.02 & 13.83$\pm$0.02 & 13.44$\pm$0.06 & 13.03$\pm$0.06 & 10.20$\pm$0.15 &  -1.26  \\ 
37 & WFI~J08415780-5252140	  & M	    & 10.18$\pm$0.01 & 10.22$\pm$0.01 & 10.17$\pm$0.01 & 10.10$\pm$0.01 &  9.73$\pm$0.15 &  -2.71  \\ 
\noalign{\medskip}
\hline
\end{tabular}
\end{center}
$^\wedge$ \footnotesize{M = confirmed cluster member; HR-C = member candidate according to the location on the HR diagram; 
X = X-ray sources \citep{Mar05}; IR-C: young stellar object candidate according to the IR properties (Sect.~\ref{sel_debris}).}\\
$^\dag$ \footnotesize{Flux at 5.8 and 8~$\mu$m not available.}\\
\end{table*}

\begin{figure} 
\resizebox{\hsize}{!}{\includegraphics[width=2.5cm,height=1.75cm]{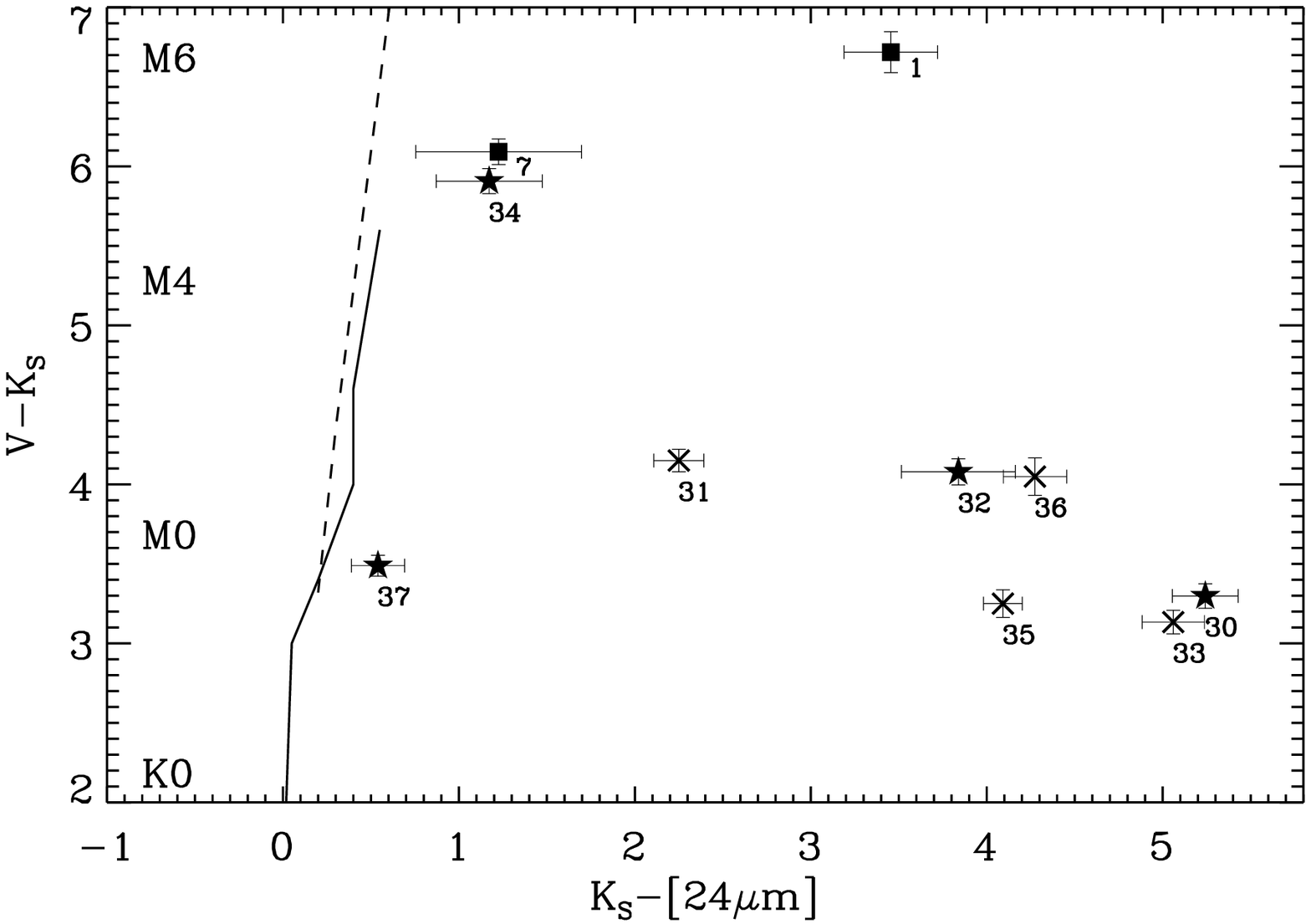}}
\caption{($V-K_S$) vs. ($K_S-[24 \mu m]$) color-color diagram for the 10
sources in IC~2391 showing IR excess at 24$\mu$m. The continuous and dashed lines are the mean 
dereddened photospheric loci determined by \citet{Sie07} and \citet{Gau07}, respectively. 
Sources redder than this threshold possess ($K_S-[24 \mu m]$) flux ratios in excess of 
expected photospheric colors. Symbols are as follows: the stars 
mark the confirmed cluster members, while squares and 
crosses mark the member candidates and the four sources of dubious nature 
discussed in Sect.~\ref{note_obj}, respectively.}
\label{fig_sel_debris}
\end{figure}
\begin{figure*} 
\centering
\includegraphics[width=11cm,height=15cm]{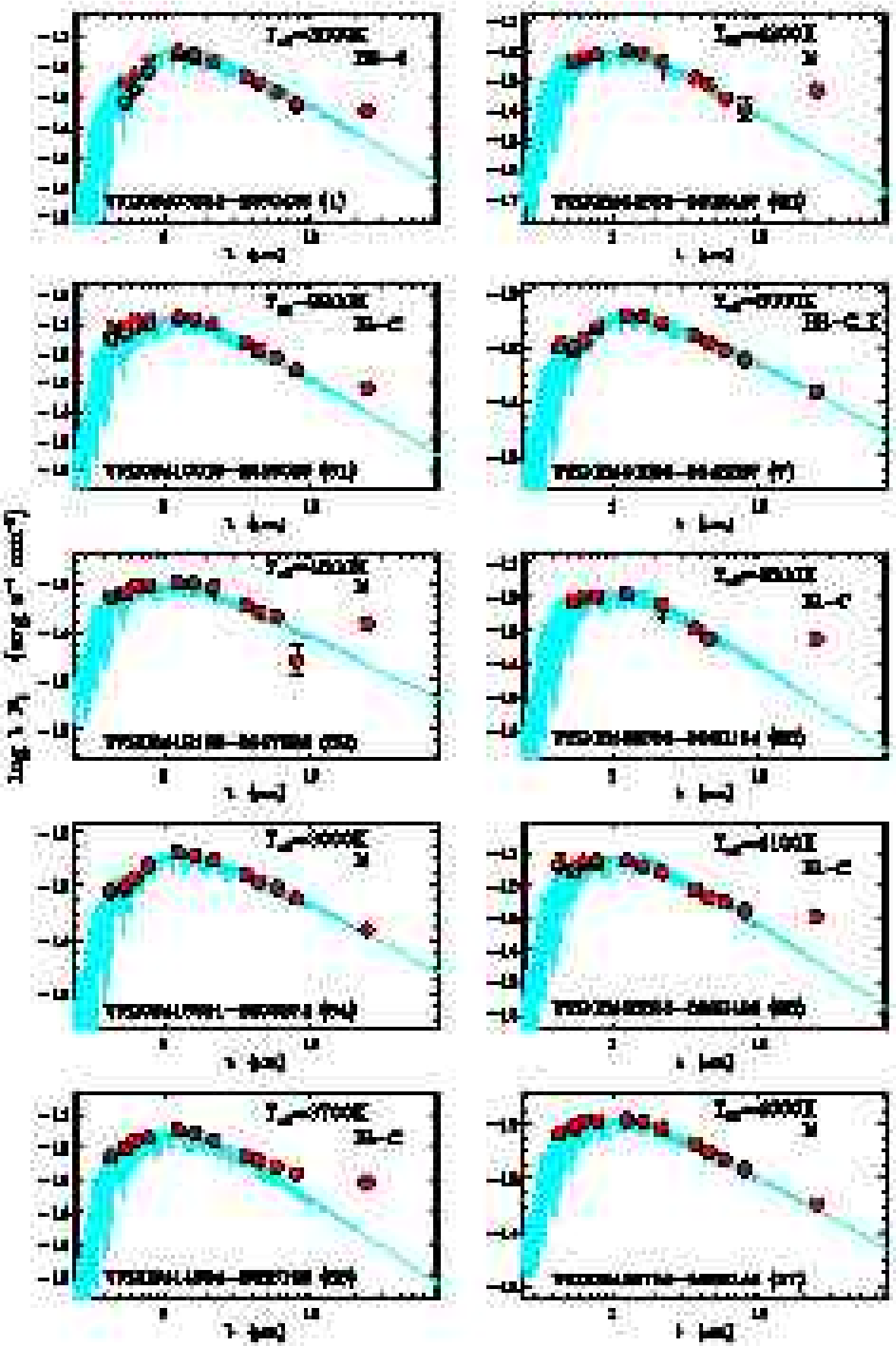}
\caption{Spectral energy distributions of the 10 sources in IC~2391 showing IR 
excess emission identified in this work.
The dereddened fluxes are represented with filled circles, while the observed 
fluxes are represented with open circles. Flux upper limits are marked with an arrow. 
The best-fit NextGen spectrum by \citet{Hau99}, for objects with T$_{\rm eff} >$4000~K, 
or StarDusty spectrum by \citet{All00}, for objects with T$_{\rm eff} \leq$4000~K, 
is overplotted on each SED, representing the stellar emission. 
Each panel is labeled with the object name (and ID number between brackets), 
estimated effective temperature (Table~\ref{tab_par}) and membership status (see Table~\ref{tab_spitzer}).}
\label{fig_seds}
\end{figure*}

We identified possible debris disk objects in IC~2391 by measuring 24$\mu$m flux 
densities in excess with respect to the expected photospheric emission; 
this criterion has been adopted in the pioneer study by \citet{Sie07} and 
allows us to pick-up new debris disk candidates in a homogeneous way. 
We use the ($V-K_S$) vs. ($K_S-[24 \mu m]$) color-color (CC) 
diagram to identify these excess sources across a broad
range of spectral types. As demonstrated by \citet{Sie07}, 
the ($V-K_S$) color is a good tracer for spectral types down to late M-dwarfs, while 
the ($K_S-[24 \mu m]$) color is a good diagnostic for mid-IR excess 
because the $K_S$-band flux is essentially photospheric for stars older than $\sim$10~Myr, 
which have already dissipated the innermost region of their disks \citep{Hai01}. 
We dereddened the magnitudes of all the sources detected at 24$\mu$m using the A$_V$ 
values derived in Sect.~\ref{par} and the extinction law by 
\citet{Car89}\footnote{We adopt the \citet{Car89} extinction law for 
consistency with our parametrization procedure (Sect.~\ref{par}).}. 
We select objects with IR excess at 24$\mu$m using the mean dereddened 
photospheric locus in the ($V-K_S$) vs. ($K_S-[24 \mu m]$) 
diagram determined by \citet{Sie07}; 
for spectral types later than M4, we use the  IR photospheric 
colors for M-dwarfs determined by \citet{Gau07}. 

As shown in Fig.~\ref{fig_sel_debris}, we found 10 sources redder than this threshold, 
i.e. objects possessing ($K_S-[24 \mu m]$) flux ratios in excess of expected 
photospheric colors by at least 15\%, i.e. the same criterion as in \citet{Sie07}. 
Four out of these 10 sources (ID~30, ID~32, ID~34 and ID~37), 
are previously known members of the cluster\footnote{See Table~\ref{tab_phot} 
for previous bibliography on these objects.} (stars in Fig.~\ref{fig_sel_debris}), 
two (ID~1 and ID~7) are new candidate members selected in this work 
(squares in Fig.~\ref{fig_sel_debris}), 
while the remaining four (ID~31, ID~33, ID~35 and ID~36) are new sources with 
IR excess emission neither reported by previous 
studies nor selected as candidate cluster members 
by our optical criteria (crosses in Fig.~\ref{fig_sel_debris}); 
the latter objects are further discussed in Sect.~\ref{note_obj}. 

The optical and near-IR photometry and the physical parameters of the 
10 sources showing IR excess at MIPS/24$\mu$m are reported in Table~\ref{tab_phot} 
and Table~\ref{tab_par}, respectively, while Table~\ref{tab_spitzer} 
reports their IRAC and MIPS/24$\mu$m magnitudes.

\subsubsection{Comments on individual objects \label{note_obj}}

\citet{Sie07} reported IR excesses at 24$\mu$m indicative of debris disks for 8 objects in IC~2391. 
Only two of these objects fall in the area observed with WFI, namely ID~29 and 
ID~24 from their Table~2; we correctly recovered ID~29 as a possible debris disk object 
(i.e. our ID~37), while ID~24 is saturated in our WFI images 
and was then \emph{a priori} excluded from our analysis.

Four of our debris disk candidates, namely ID~31, ID~33, ID~35 and ID~36, though showing
a clear IR excess emission in their SEDs (Fig.~\ref{fig_seds}), were not selected as 
candidate cluster members neither by previous surveys nor by our optical criteria. 
This is because these four sources are sub-luminous in the HR diagram when 
adopting the distance of 150~pc and the extinction determined as explained in Sect.~\ref{par}. 
Possible reasons for this inconsistency may be an underestimate of the extinction 
and/or the distance for these four sources. 
From a visual inspection of our optical WFI images, these four sources appear point-like 
and are concentrated in the central area of the cluster (Fig.~\ref{fig:obs}); 
moreover, by fitting their SEDs with the \citet{Rob06} grid of models for young stellar objects 
normalized to the observed $I$-band flux, 
we obtain stellar and disk parameters still consistent with young stars 
belonging to the cluster (see Table~\ref{tab_par}). 
Thus, for the time being, we include these sources in our list of 
possible debris disk objects in IC~2391; 
future follow-up spectroscopy will shed light on their nature.

\subsection{On the disk properties \label{pardisk}}

This section aims for a description of the circumstellar material 
around the 10 objects with IR excess emission selected above, in order 
to give some hint on the nature and evolutionary phase of their disks.
Recent publications \citep[e.g.][]{Har07,Alc08,Mer08} of Spitzer observations in star forming regions 
within the frame of the c2d survey have provided several methods to study disk properties such as the 
amount of circumstellar dust, morphology, etc. 
We applied some of these methods to further investigate 
the nature of our 10 debris disk candidates in IC~2391.

We first construct the SEDs of these objects from optical to mid-IR wavelengths, 
as already described in Sect.~\ref{par} but including IRAC and MIPS data and, where possible, 
additional $B$ magnitudes from the NOMAD catalog \citep{Zac05}. 
For each object the flux at each wavelength was corrected for interstellar extinction using 
the A$_V$ values reported in Table~\ref{tab_par} and the normal (R$_V$=3.1) reddening
law by \citet{Car89}. The SEDs of the 10 debris disk candidates in IC~2391 
are shown in Fig.~\ref{fig_seds} together with the best-fitting model spectra 
by \citet{Hau99} and \citet{All00} with the same temperature 
as the objects (see Sect.~\ref{par}), representing the stellar contribution. 
For the majority of the objects the IR excess with respect to 
the expected photospheric emission is clearly visible. Moreover, the 
lack of near-IR excess emission suggests the presence of disks whose 
inner part has been already dissipated \citep[see, e.g.,][]{Sic06}.

\begin{figure} 
\resizebox{\hsize}{!}{\includegraphics[width=2.5cm,height=2cm]{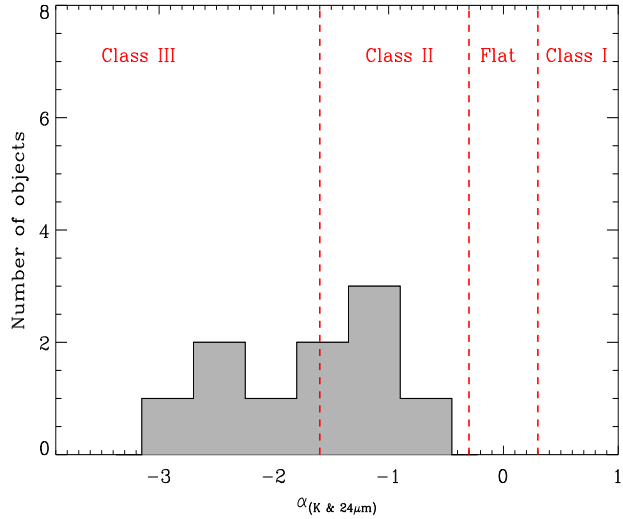}}
\caption{Distribution of the $\alpha_{[K \& 24\mu m]}$ SED slopes 
for the sources in IC~2391 showing IR excess emission identified in this work. 
The dashed vertical lines mark the boundaries of class I, flat, class II and class III objects according 
to the classification by \citet{Lad87} as extended by \citet{Gre94}}.
\label{histo_Ldisk}
\end{figure}

Additional clues on the nature of our debris disk candidates can be 
obtained by investigating the objects IR classes according 
to the definition by \citet{Lad87} as extended by \citet{Gre94}. 
This classification is traditionally based on the value of the $\alpha$ spectral index, 
which is determined as the slope of the SED in the form 
$\log \lambda F_{\lambda}$ vs. $\log \lambda$ at wavelengths 
longer than 2$\mu$m. Following the c2d convention \citep[e.g.][]{Har07, Alc08, Mer08}, 
we use $\alpha_{[K \& 24\mu m]}$ as the reference spectral index, 
which is the slope of the linear fit to the fluxes at
$K_S$ and MIPS/24$\mu$m (Table~\ref{tab_spitzer}). 
As shown in Fig.~\ref{histo_Ldisk}, 
we found that 5 out of the 10 debris disk candidates in our sample present 
$\alpha_{[K \& 24\mu m]}$ slopes typical of IR class III sources ($\alpha_{[K \& 24\mu m]}<$-1.6) 
according to the classification by \citet{Gre94}, i.e. they have optically thin disks. 
The remaining 5 objects fall in the class II regime (-1.6$<\alpha_{[K \& 24\mu m]} \leq$-0.3), 
i.e. they would have optically thick disks; however, the majority of them lay very 
close to the class III boundary (i.e. $\alpha_{[K \& 24\mu m]} \lesssim$-1). 
The possibility that IC~2391 would have any thick/transitional disks is quite surprising 
considering its age (30-50~Myr), because these disks, along with primordial disks, are  
typically dissipated within $\sim$10~Myr \citep{Dec03,Dom03}. 
However, Spitzer data have shown a very clear mass dependence of
the disk dissipation timescale, i.e. very low-mass stars dissipate 
their primordial disks slower than more massive stars \citep{Sch07,Hil08}.

In conclusion, the 10 sources with IR excess emission selected in 
Sect.~\ref{sel_debris} are likely to possess evolved disks, 
whose inner part might have already been cleared by the formation 
of large-sized bodies (such as rocks). 


\section{Conclusions \label{conclu}}

We performed a deep wide-field imaging survey of the young open cluster 
IC~2391 to study its low mass population down to 0.03~M$_{\odot}$. 
We used our optical $VRI$ data, in combination with $JHK_S$ magnitudes 
from the 2MASS catalog to estimate the physical parameters of the 
optically detected sources and selected cluster member candidates 
on the basis of their location on the HR diagram compared to theoretical 
isochrones. We estimated the contamination using 
statistical arguments and, when possible, proper motion from the NOMAD catalog. 
Our survey has a completeness level of 80\% down to $I_C$=23.0, i.e. 
$\sim$1-2~mag fainter than previous optical surveys conducted in IC~2391, and 
revealed 29 new low-mass member candidates, 
among which 12 may be BDs. 
The expected contaminations is at least $\sim$50\%. 

We confirm the presence of a significant drop of the cluster mass spectrum 
across the stellar/substellar boundary, with the fraction of sub-stellar 
objects being of only 8-15\%. 
The lack of BDs in IC~2391 might be due to the 
dip in the luminosity function around M7-M8 already observed in 
other star-forming regions, open clusters 
and the field, which is caused by the beginning of larger-size dust 
formation in the atmospheres of objects in this effective temperature regime. 
However, other explanations such as dynamical mass segregation and/or 
masking in binary systems might be invoked. 
Our data are not sufficient to decide between these scenarios.

Finally, we have combined our ground-based optical observations with near-IR 
photometry from 2MASS and Spitzer GTO observations to investigate 
the debris disk population of IC~2391. 
We identified 10 possible debris disk objects in the cluster on the basis of 
their 24$\mu$m flux densities in excess with respect to the expected 
photospheric emission. We constructed optical/IR SEDs for all these 
objects and investigate the properties of their circumstellar material. 
According to our analysis, these 10 objects are likely to possess 
evolved disks whose inner part has been already cleared up. 
A few of these objects might possess thick/transition disks, 
what would be surprising, given the age of IC~2391. 

\begin{acknowledgements}

This paper is based on observations carried out at the European 
Southern Observatory, La Silla (Chile), under observing program number 68.D-0541(A). 
This publication makes use of data products from the Two Micron All Sky
Survey, which is a joint project of the University of Massachusetts and
the Infrared Processing and Analysis Center/California Institute of
Technology, funded by NASA and the National Science Foundation. 
We also acknowledge extensive use of the SIMBAD database, 
operated at CDS Strasbourg, the WEBDA database, operated at the Institute for 
Astronomy of the University of Vienna, and the NOMAD catalog, 
released by the U.S. Naval Observatory. 
L. Spezzi acknowledges financial support from INAF-Catania. 
We thank J.M. Alcal\'a, A. Frasca, D. Gandolfi and F. Comer\'on 
for many discussions and suggestions. 
We are also grateful to many others, in particular to Salvatore Spezzi.

\end{acknowledgements}

\bibliographystyle{aa}

\end{document}